\documentclass[journal]{IEEEtran}

\usepackage{amsmath}
\usepackage{amsthm}
\usepackage{amssymb}

\usepackage{subcaption}
\usepackage[font=small,labelfont=bf]{caption}

\usepackage{url}
\usepackage{cite}
\usepackage{graphicx}
\usepackage{color}
\usepackage[percent]{overpic}

\newtheorem{lemma}{Lemma}
\newtheorem{observation}{Observation}

\begin{document}

\newcommand{\mathacr}[1]{\mathsf{#1}}
\newcommand{\argmax}[1]{{\underset{{#1}}{\mathrm{arg\,max}}}}
\newcommand{\argmin}[1]{{\underset{{#1}}{\mathrm{arg\,min}}}}
\newcommand{\vect}[1]{\mathbf{#1}}
\newcommand{\maximize}[1]{{\underset{{#1}}{\mathrm{maximize}}}}
\newcommand{\minimize}[1]{{\underset{{#1}}{\mathrm{minimize}}}}
\newcommand{\condSum}[3]{\overset{#3}{\underset{\underset{#2}{#1}}{\sum}}}
\newcommand{\condProd}[3]{\overset{#3}{\underset{\underset{#2}{#1}}{\prod}}}

\def\diag{\mathrm{diag}}
\def\kron{\otimes}
\def\tr{\mathrm{tr}}
\def\rank{\mathrm{rank}}
\def\Htran{\mbox{\tiny $\mathrm{H}$}}
\def\Ttran{\mbox{\tiny $\mathrm{T}$}}
\def\imagunit{\mathsf{j}} 
\def\CN{\mathcal{N}_{\mathbb{C}}} 
\def\EcondH{\mathbb{E}_{| \vect{H}}}
\def\boff{b_{\mathrm{off}}}
\def\erf{\mathrm{erf}}

%
% paper title

\title{Hardware Distortion Correlation Has Negligible Impact on UL Massive MIMO Spectral Efficiency}

\author{
\IEEEauthorblockN{Emil Bj{\"o}rnson, \emph{Senior Member, IEEE}, Luca Sanguinetti, \emph{Senior Member, IEEE}, Jakob Hoydis, \emph{Member, IEEE}
\thanks{
\copyright 2018 IEEE. Personal use of this material is permitted. Permission from IEEE must be obtained for all other uses, in any current or future media, including reprinting/republishing this material for advertising or promotional purposes, creating new collective works, for resale or redistribution to servers or lists, or reuse of any copyrighted component of this work in other works.
\newline\indent This paper was supported by ELLIIT and CENIIT.
\newline \indent E.~Bj\"ornson is with the Department of Electrical Engineering (ISY), Link\"{o}ping University, 58183 Link\"{o}ping, Sweden (emil.bjornson@liu.se). \newline \indent L.~Sanguinetti is with the University of Pisa, Dipartimento di Ingegneria dell'Informazione, 56122 Pisa, Italy (luca.sanguinetti@unipi.it). 
\newline \indent
J.~Hoydis is with Nokia Bell Labs, Paris-Saclay, 91620 Nozay, France (jakob.hoydis@nokia-bell-labs.com).
\newline\indent Parts of this paper was presented at the IEEE International Workshop on Signal Processing Advances in Wireless Communications (SPAWC), 2018 \cite{Bjornson2018a}.}
% make the title area
}}
\maketitle

% The paper headers
\markboth{IEEE Transactions on Communications}%
{Shell \MakeLowercase{\textit{et al.}}: Bare Demo of IEEEtran.cls for IEEE Journals}

% make the title area
\maketitle

\begin{abstract}
This paper analyzes how the distortion created by hardware impairments in a multiple-antenna base station affects the uplink spectral efficiency (SE), with focus on Massive MIMO. This distortion is correlated across the antennas, but has been often approximated as uncorrelated to facilitate (tractable) SE analysis. To determine when this approximation is accurate, basic properties of distortion correlation are first uncovered. Then, we separately analyze the distortion correlation caused by third-order non-linearities and by quantization. Finally, we study the SE numerically and show that the distortion correlation can be safely neglected in Massive MIMO when there are sufficiently many users. Under i.i.d.~Rayleigh fading and equal signal-to-noise ratios (SNRs), this occurs for more than five transmitting users. Other channel models and SNR variations have only minor impact on the accuracy. We also demonstrate the importance of taking the distortion characteristics into account in the receive combining.
\end{abstract}


\IEEEpeerreviewmaketitle

\section{Introduction}

The common practice in academic communication research is to consider ideal transceiver hardware whose impact on the transmitted and received signals is negligible.
However, the receiver hardware in a wireless communication system is always non-ideal \cite{Schenk2008a}; for example, there are non-linear amplifications in the low-noise amplifier (LNA) and quantization errors in the analog-to-digital converter (ADC). Hence, there is a mismatch between the signal that reaches the antenna input and the signal that is obtained in the digital baseband.
In single-antenna receivers, the mismatch over a frequency-flat channel can be equivalently represented by scaling the received signal and adding uncorrelated (but not independent) noise, using the Bussgang theorem \cite{Bussgang1952a,Fletcher2007a}.
This is a convenient representation in information theory, since one can compute lower bounds on the capacity by utilizing the fact that the worst-case uncorrelated additive noise is independent and Gaussian distributed~\cite{Hassibi2003a}.

The extension of the Bussgang theorem to multi-antenna receivers is non-trivial, but important since Massive MIMO (multiple-input multiple-output) is the key technology to improve the spectral efficiency (SE) in future networks \cite{Parkvall2017a, Larsson2014a, massivemimobook}. In Massive MIMO, a base station (BS) with $M$ antennas is used to serve $K$ user equipments (UEs) by spatial multiplexing \cite{Marzetta2010a, Larsson2014a,massivemimobook}. 
By coherently combining the signals that are received and transmitted from the BS's antenna array, the SE can be improved by both an array gain and a multiplexing gain; the former improves the per-user SE (particularly for cell-edge UEs) and the latter can lead to orders-of-magnitude improvements in per-cell SE \cite{Bjornson2016a}.
While the achievable per-cell SE was initially believed to be upper limited by pilot contamination \cite{Marzetta2010a}, it has recently been shown that the spatial channel correlation that is present in any practical channel can be exploited to alleviate that effect \cite{ashikhmin2012pilot, Yin2013a, BHS18A}. The typical Massive MIMO configuration is $M \approx 100$ antennas and $K\in[10,50]$ UEs \cite[Sec.~7.2]{massivemimobook}, \cite{Shepard2013a}, \cite{Harris2017a}, which is ``massive'' as compared to conventional multiuser MIMO. These parameter ranges are what we refer to as Massive MIMO in this paper.

In the uplink, each  BS antenna is equipped with a separate transceiver chain (including LNA and ADC), so that one naturally assumes the distortion is uncorrelated between antennas. This assumption was made in \cite{Bai2013a,Orhan2015a} (among others) without discussion, while the existence of correlation was mentioned in \cite{Bjornson2014a,Jacobsson2017b}, but claimed to be negligible for the SE and energy efficiency analysis of Massive MIMO systems under Rayleigh fading.
However, if two antennas equipped with identical hardware would receive exactly the same signal, the distortion would clearly be identical. Hence, it is important to characterize under which conditions, if any, the distortion can be reasonably modeled as uncorrelated across the receive antennas. 

The distortion correlation is studied both analytically and experimentally in \cite{Moghadam2012a}, in the related scenario of multi-antenna transmission with non-linear hardware. The authors observed distortion correlation for single-stream transmission in the sense that each element of the complex-valued distortion vector has the same angle as the corresponding element in the signal vector, leading to a similar directivity of the radiated signal and distortion.
However, the authors of \cite{Moghadam2012a} conjectured that the hardware distortion is ``\emph{practically uncorrelated}'' when transmitting multiple precoded data streams. In that case, each signal vector is a linear combination of the precoding vectors where the coefficients are the random data. Hence, the directivity changes rapidly with the data and does not match any of the individual precoding vectors.
The conjecture from \cite{Moghadam2012a} was proved analytically in \cite{Mollen2018b} for the case when many data streams are transmitted with similar power. The error-vector magnitude analysis in \cite{Gustavsson2014a} shows that an approximate model where the BS distortion is uncorrelated across the antennas gives accurate results, while 
\cite{Handel2018a} provides a similar result when considering a model that also includes mutual coupling and noise at the transmitter side.
 Nevertheless, \cite{Zou2015a} recently claimed that the model with uncorrelated BS distortion ``\emph{do not necessarily reflect the true behavior of the transmitted distorted signals}'', \cite{Larsson2018a} stated that the model is ``{\emph{physically inaccurate}}'', and \cite{Mollen2018a} says that it  ``{\emph{can yield incorrect conclusions in some important cases of interest when the number of antennas is large}}''. These papers focus on the power spectral density and out-of-band interference, assuming setups with non-linearities at the BS, while the UEs are equipped with ideal hardware. In contrast, the works \cite{Bjornson2014a,massivemimobook} promoting the uncorrelated BS distortion model also consider hardware distortion at the UEs and claim that it is the dominant factor in Massive MIMO. Hence, the question is if the statements from \cite{Zou2015a,Larsson2018a,Mollen2018a} are applicable when quantifying the uplink SE in typical Massive MIMO scenarios with hardware impairments at both the UEs and BSs. In other words:

\begin{center}
Is the SE analysis in prior works on Massive MIMO with uncorrelated BS distortion trustworthy?
\end{center}

\subsection{Main Contributions}

This paper takes a closer look at the question above, with the aim of providing a versatile view of when models with uncorrelated hardware distortion can be used for SE analysis in the uplink with multiple-antenna BSs. To this end, we develop in Section~\ref{sec:system-model} a system model with a non-ideal multiple-antenna receiver and explain the characteristics of hardware distortion along the way. 
Achievable SE expressions are provided and we develop a new receive combining scheme that maximizes the SE. The analysis is performed under the assumption of perfect channel state information (CSI), while the imperfect case is studied numerically.
We study amplitude-to-amplitude (AM-AM) non-linearities analytically in Section~\ref{sec:non-linearities}, with focus on maximum ratio (MR) combining. 
The characteristics of the quantization distortion are then analyzed in Section~\ref{sec:quantization}. 
We quantify the SE with both non-linearities and quantization errors in Section~\ref{sec:SE-analysis}. In particular, we compare the SE with distortion correlation and the corresponding SE under the simplifying assumption (approximation) of uncorrelated distortion, to show under which conditions these are approximately equal. The major conclusions are then summarized in Section~\ref{sec:conclusion}, while the extensive set of conclusions are clearly marked throughout the paper.

We consider a single-cell scenario with $K$ UEs in this paper, for notational convenience. However, the analytical results are readily extended to a multi-cell scenario by letting some of the $K$ UEs represent UEs in other cells. The only difference is that we should only compute the SE of the UEs that reside in the cell under study.

The conference version \cite{Bjornson2018a} of this paper considers only non-linearities and contains a subset of the analysis and proofs.

\textit{\textbf{Reproducible research:}} All the simulation results can be reproduced using the Matlab code and data files available at:\\ \url{https://github.com/emilbjornson/distortion-correlation}

\textit{\textbf{Notation:}} Boldface lowercase letters, $\vect{x}$, denote column vectors and boldface uppercase letters, $\vect{X}$, denote matrices. The $n \times n$ identity matrix is $\vect{I}_n$. The superscripts $^{\Ttran}$, $^\star$ and $^{\Htran}$ denote transpose, conjugate, and conjugate transpose, respectively. The Moore-Penrose inverse is denoted by $^{\dagger}$, 
$\odot$ denotes the Hadamard (element-wise) product of matrices, $\erf(\cdot)$ is the error function, $\imagunit$ is the imaginary unit, while $\Re(\cdot)$ and $\Im(\cdot)$ give the real and imaginary part of a complex number, respectively.
The multi-variate circularly symmetric complex Gaussian distribution with correlation matrix $\vect{R}$ is denoted as $\CN(\vect{0},\vect{R})$. The expected value of a random variable $x$ is $\mathbb{E}\{ x \}$.

\section{System Model and Spectral Efficiency}
\label{sec:system-model}

We consider the uplink of a single-cell system where $K$ single-antenna UEs communicate with a BS equipped with $M$ antennas. We consider a symbol-sampled complex baseband system model \cite{Mollen2018a}.\footnote{This model is sufficient to study the inband communication performance, while a continuous-time model is needed to capture out-of-band interference. Hence, we are implicitly assuming that no out-of-band interference exists. See  \cite{Larsson2018a,Mollen2018a} for recent studies of out-of-band distortion in Massive MIMO.}
 The channel from UE $k$ is denoted by $\vect{h}_k =[h_{k1} \, \ldots \, h_{kM}]^{\Ttran} \in \mathbb{C}^{M}$. A block-fading model is considered where the channels are fixed within a time-frequency coherence block and take independent realizations in each block, according to an ergodic stationary  random process. In an arbitrary coherence block, the noise-free signal $\vect{u} = [u_1 \, \ldots \, u_M]^{\Ttran} \in \mathbb{C}^{M}$ received at the BS antennas is
\begin{equation}
\vect{u} = \sum_{k=1}^{K} \vect{h}_k s_k = \vect{H} \vect{s}
\end{equation}
where $\vect{H} = [\vect{h}_1 \, \ldots \, \vect{h}_{K}]\in \mathbb{C}^{M\times K}$ is the channel matrix and $s_k \in \mathbb{C}$ is the information-bearing signal transmitted by UE $k$. Since we want to quantify the SE and Gaussian codebooks maximizes the differential entropy, we assume $\vect{s}= [s_1 \, \ldots \, s_M]^{\Ttran} \sim \CN(\vect{0},p \vect{I}_K)$ where $p$ is the variance. The UEs can have unequal pathlosses and/or transmit powers, which are both absorbed into how the channel realizations $\vect{h}_k$ are generated. Several different channel distributions are considered in Section~\ref{sec:SE-analysis}.
We will now analyze how $\vect{u}$ is affected by non-ideal receiver hardware and additive noise. To focus on the distortion characteristics only, $\vect{H}$ is assumed to be known in the analytical part of the paper. Imperfect channel knowledge is considered in Section~\ref{subsec:imperfect_CSI}.

\subsection{Basic Modeling of BS Receiver Hardware Impairments} \label{subsec_basic_modeling}

We focus now on an arbitrary coherence block with a fixed channel realization $\vect{H}$ and use $\EcondH\{ \cdot \}$ to denote the conditional expectation given $\vect{H}$. Hence, the conditional distribution of $\vect{u}$ is $\CN(\vect{0},\vect{C}_{uu})$ where $\vect{C}_{uu} = \EcondH\{\vect{u} \vect{u} ^{\Htran}\} = p \vect{H} \vect{H}^{\Htran} \in \mathbb{C}^{M \times M}$ describes the correlation between signals received at different antennas. It is only when $\vect{C}_{uu}$ is a scaled identity matrix that $\vect{u}$ has uncorrelated elements. This can only happen when $K \geq M$. However, we stress that $K < M$ is of main interest in Massive MIMO \cite{massivemimobook} (or more generally in any multiuser MIMO system).

\begin{figure*}
  \centering 
  		\begin{overpic}[width=0.5\textwidth,tics=10]{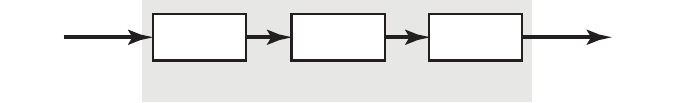}
 \put (4,8) {$\vect{u}$}
  \put (24.5,8) {LNA}
    \put (46,8) {I/Q}
      \put (64.2,8) {ADC}
      \put (26,1) {Non-ideal BS receiver}  
  \put (91,8) {$\boldsymbol{g}(\vect{u}) = \vect{D}\vect{u} + \boldsymbol{\eta}$}
    \end{overpic}    
      \caption{The signal vector $\vect{u}$ that reaches the antennas is processed by the non-ideal BS receiver hardware, which includes LNA, I/Q demodulator, and ADC. As proved in Section~\ref{subsec_basic_modeling}, the output signal $\boldsymbol{g}(\vect{u})$, which is fed to the digital baseband processor, can be equivalently expressed as $ \vect{D}\vect{u} + \boldsymbol{\eta}$, where the additive distortion vector $\boldsymbol{\eta}$ has correlated elements but is uncorrelated with $\vect{u}$.} \label{figure_setup}   \vspace{-3mm}
\end{figure*}
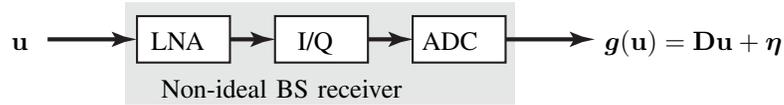

The BS hardware is assumed to be non-ideal but quasi-memoryless, which means that both the amplitude and phase are affected \cite{Raich2002a}. The impairments at antenna $m$ are modeled by an arbitrary deterministic function $g_m(\cdot): \mathbb{C} \to \mathbb{C}$, for $m=1,\ldots,M$, which can describe both continuous non-linearities and discontinuous quantization errors.
These functions distort each of the components in $\vect{u}$, such that the resulting signal is
\begin{equation} \label{eq:z-distortion}
\vect{z} = \begin{bmatrix}
g_1(u_1) \\ \vdots \\ g_M(u_M)
\end{bmatrix} \triangleq \boldsymbol{g}(\vect{u}).
\end{equation}
By defining $\vect{C}_{zu} = \EcondH\{ \vect{z} \vect{u}^{\Htran} \}$, we can express $\vect{z}$ as a linear function of the input $\vect{u}$ by exploiting that $\vect{C}_{zu} \vect{C}_{uu}^{\dagger} \vect{u}$ is the linear minimum mean-squared error (LMMSE) estimate of $\vect{z}$ given $\vect{u}$ \cite[Sect.~15.8]{Kay1993a}. Notice that we used the Moore-Penrose inverse since $\vect{C}_{uu}$ is generally rank-deficient. We can further define the ``estimation error'' $\boldsymbol{\eta} \triangleq \boldsymbol{g}(\vect{u}) - \vect{C}_{zu} \vect{C}_{uu}^{\dagger} \vect{u}$, which we will call the \emph{additive distortion term}. This leads to the input-output relation
\begin{equation} \label{eq:z-distortion2}
\vect{z} = \boldsymbol{g}(\vect{u}) = \vect{C}_{zu} \vect{C}_{uu}^{\dagger} \vect{u} + \boldsymbol{\eta}.
\end{equation}
By construction, the signal $\vect{u} $ is uncorrelated with $\boldsymbol{\eta}$; that is,\footnote{This result holds even if $\vect{C}_{uu} = p \vect{H}\vect{H}^{\Htran}$ is rank-deficient, in which case the Moore-Penrose inverse can be obtained by computing an eigenvalue decomposition of $\vect{C}_{uu}$ and inverting all the non-zero eigenvalues.
To prove the second equality, we need to utilize the fact that the null space of 
$\vect{C}_{uu}$ is contained in the null space of $\vect{C}_{zu} = \EcondH\{ \vect{z} \vect{s}^{\Htran}  \} \vect{H}^{\Htran}$.}
\begin{equation}
\EcondH \{ \boldsymbol{\eta} \vect{u}^{\Htran} \} = \vect{C}_{zu} -  \vect{C}_{zu} \vect{C}_{uu}^{\dagger} \vect{C}_{uu} = \vect{0}.
\end{equation}
However, $\vect{u} $ and $\boldsymbol{\eta}$ are clearly not independent.

This derivation has not utilized the fact that $\vect{u}$ is Gaussian distributed (for a given $\vect{H}$), but only its first and second order moments. Hence, the model in \eqref{eq:z-distortion2} holds also for finite-sized constellations. Since the SE will be our metric, we utilize the full distribution to simplify $\vect{C}_{zu} \vect{C}_{uu}^{\dagger}$ in \eqref{eq:z-distortion2} using a discrete complex-valued counterpart to the Bussgang theorem \cite{Bussgang1952a}.

\begin{lemma} \label{lemma:bussgang}
Consider the jointly circular-symmetric complex Gaussian variables $x$ and $y$. For any deterministic function $f(\cdot): \mathbb{C}\to \mathbb{C}$, it holds that
\begin{equation} \label{eq:Bussgang-expression}
\mathbb{E}\{ f(x) y^*\} = \mathbb{E}\{ f(x) x^*\} \frac{\mathbb{E}\{ x y^*\}}{\mathbb{E}\{ |x|^2 \}}.
\end{equation}
\end{lemma}
\begin{IEEEproof}
Note that $y = \frac{\mathbb{E}\{ y x^*\}}{\mathbb{E}\{ |x|^2 \}} x + \epsilon$, where $\epsilon = y- \frac{\mathbb{E}\{ y x^*\}}{\mathbb{E}\{ |x|^2 \}} x$ has zero mean and is uncorrelated with $x$. The Gaussian distribution implies that $\epsilon$ and $x$ are independent. Replacing $y$ with this expression in the left-hand side of \eqref{eq:Bussgang-expression} and noting that $\mathbb{E}\{ f(x) \epsilon^*\} =0$ yields the right-hand side of \eqref{eq:Bussgang-expression}.
\end{IEEEproof}

Using Lemma~\ref{lemma:bussgang}, it follows that (cf.~\cite[Appendix~A]{Jacobsson2017a})
\begin{equation} \label{eq:Czu_expression}
\vect{C}_{zu} =\EcondH\{ \vect{z} \vect{u}^{\Htran} \}= \vect{D} \vect{C}_{uu}
\end{equation}
where $\vect{D} = \diag(d_1,\ldots,d_M)$ and $d_m = \frac{\EcondH\{ g_m(u_m) u_m^\star \}}{  \EcondH \{ | u_m |^2 \}}$. 
Inserting \eqref{eq:Czu_expression} into  \eqref{eq:z-distortion2} yields
\begin{equation} \label{eq:z-distortion3}
\vect{z} = \vect{D}\vect{u} + \boldsymbol{\eta}
\end{equation}
with $\boldsymbol{\eta} = \boldsymbol{g}(\vect{u}) - \vect{D}\vect{u}$. The following observation 
is thus made.

\begin{observation}
When a Gaussian signal $\vect{u}$ is affected by non-ideal BS receiver hardware, the output is an element-wise scaled version of $\vect{u}$ plus a distortion term $\boldsymbol{\eta}$ that is uncorrelated (but not independent) with $\vect{u}$.
\end{observation}

If the functions $g_m(\cdot)$ are all equal and $\vect{C}_{uu}$ has identical diagonal elements, $\vect{D}$ is a scaled identity matrix and the common scaling factor represents the power-loss jointly incurred by the hardware impairments. In this special case, $ \vect{D}\vect{u}$ has the same correlation characteristics as $\vect{u}$.

The distortion term has the (conditional) correlation matrix
\begin{equation} \label{eq:Cetaeta}
\vect{C}_{\eta \eta} = \EcondH \{\boldsymbol{\eta}\boldsymbol{\eta}^{\Htran}\} = \vect{C}_{zz} - \vect{C}_{zu} \vect{C}_{uu}^{\dagger} \vect{C}_{zu}^{\Htran} =  \vect{C}_{zz} - \vect{D} \vect{C}_{uu} \vect{D}^{\Htran}
\end{equation}
where $ \vect{C}_{zz} = \EcondH \{ \vect{z} \vect{z}^{\Htran} \}$. In the special case when $\vect{C}_{uu}$ is diagonal and $\EcondH\{ g_m(u_m) \} = 0$ for $m=1,\ldots,M$, $\vect{C}_{\eta \eta}$ is also diagonal and the distortion term has uncorrelated elements. This cannot happen unless $K \geq M$.

\begin{observation}
The elements of the BS distortion term $\boldsymbol{\eta}$ are always correlated in multiuser MIMO if $K < M$.
\end{observation}

This confirms previous downlink \cite{Moghadam2012a,Zou2015a,Mollen2018b,Larsson2018a} and uplink \cite{Mollen2018a} results, derived with different system models. 
The equivalent model of the receiver distortion derived in this section is summarized in Fig.~\ref{figure_setup}.
The question is now if the simplifying assumption of uncorrelated distortion (e.g., made in \cite{Bai2013a,Orhan2015a,Bjornson2014a}) has significant impact on the SE.

\subsection{Spectral Efficiency with BS Hardware Impairments}

Using the signal and distortion characteristics derived above, we can compute the SE. The signal detection is based on the received signal $\vect{y}  \in \mathbb{C}^{M}$ that is available in the digital baseband and is given by
\begin{align} 
\vect{y} = \vect{z} + \vect{n} =  \vect{D}\vect{u} + \boldsymbol{\eta} + \vect{n}  =  \sum_{k=1}^{K}\vect{D}\vect{h}_k s_k + \boldsymbol{\eta} + \vect{n}   \label{eq:distorted-received-signal}
\end{align}
where $\vect{n} \sim \CN(\vect{0},\sigma^2 \vect{I}_M)$ accounts for thermal noise that is (conditionally) uncorrelated\footnote{In practice, the initially independent noise at the BS is also distorted by the BS hardware, which results in uncorrelated noise (conditioned on a channel realization $\vect{H}$) by following the same procedure as above.} with $\vect{u}$ and $\boldsymbol{\eta}$. The combining vector $\vect{v}_k$ is used to detect the signal of UE $k$ as
\begin{align}
\vect{v}_k^{\Htran}\vect{y} = \vect{v}_k^{\Htran} \vect{D}\vect{h}_k s_k +   \sum_{i=1, i\ne k}^{K} \vect{v}_k^{\Htran} \vect{D}\vect{h}_i s_i + \vect{v}_k^{\Htran}\boldsymbol{\eta} + \vect{v}_k^{\Htran}\vect{n}   \label{eq:distorted-received-signal2}.
\end{align}
In the given coherence block, $\boldsymbol{\eta}$ is uncorrelated with $\vect{u}$, thus the distortion is also uncorrelated with the information-bearing signals $s_1,\ldots,s_K$ (which are mutually independent by assumption).
Hence, we can use the \emph{worst-case uncorrelated additive noise theorem} \cite{Hassibi2003a} to lower bound the mutual information between the input $s_k$ and output $\vect{v}_k^{\Htran}\vect{y}$ in \eqref{eq:distorted-received-signal2} as 
\begin{align}  
\mathcal{I}(s_k; \vect{v}_k^{\Htran}\vect{y}) \geq \log_2 \left( 1 + \gamma_k \right) \label{eq:first-SE-bound}
\end{align}
for the given deterministic  channel realization $\vect{H}$,
where $\gamma_k $ represents the instantaneous signal-to-interference-and-noise ratio (SINR) and is given by
\begin{align}\label{eq:SINR}
\gamma_k = \frac{ p \vect{v}_k^{\Htran} \vect{D}\vect{h}_k \vect{h}_k^{\Htran} \vect{D}^{\Htran}  \vect{v}_k}{  \vect{v}_k^{\Htran} \big(\sum\limits_{i \neq k} p \vect{D}\vect{h}_i \vect{h}_i^{\Htran} \vect{D}^{\Htran} + \vect{C}_{\eta \eta}  +  \sigma^2 \vect{I}_M \big)  \vect{v}_k}.
\end{align}
The lower bound represents the worst-case situation when the uncorrelated distortion plus noise term $\boldsymbol{\eta}+\vect{n}$ is colored Gaussian noise that is independent of the desired signal and distributed as $\boldsymbol{\eta}+\vect{n} \sim \CN (\vect{0}, \vect{C}_{\eta \eta} +\sigma^2 \vect{I}_M)$. Note that this is only the worst-case conditional distribution in a coherence block (for a given $\vect{H}$), while the marginal distribution of $\boldsymbol{\eta}+\vect{n}$ is the product of Gaussian random variables \cite[Sec.~6.1]{massivemimobook}.

\begin{observation}
Treating the BS distortion as statistically independent of the desired signal is a worst-case modeling when computing the mutual information.
\end{observation}

Since $\gamma_k$ is a generalized Rayleigh quotient with respect to $\vect{v}_k$, it is maximized by \cite[Lemma B.10]{massivemimobook}
\begin{align} \label{eq:combining-vector}
\vect{v}_k &= p \bigg( \sum_{i=1,i \neq k}^K p \vect{D}\vect{h}_i \vect{h}_i^{\Htran} \vect{D}^{\Htran} \!+\! \vect{C}_{\eta \eta} \! +\!  \sigma^2 \vect{I}_M \bigg)^{\!-1}  \vect{D} \vect{h}_k  \\
&= p \bigg( \vect{C}_{z z} \! +\!  \sigma^2 \vect{I}_M -  \vect{D} \vect{h}_k \vect{h}_k^{\Htran}   \vect{D}^{\Htran}   \bigg)^{\!-1}  \vect{D} \vect{h}_k.
\end{align}
We call this the \emph{distortion-aware minimum-mean squared error} (DA-MMSE) receiver as it takes into account not only inter-user interference and noise, but also the distortion correlation. This new combining scheme is a novel contribution of this paper.\footnote{MMSE-like combining schemes have been utilized in prior works on hardware-impaired Massive MIMO that assumed uncorrelated BS distortion (see e.g.~\cite{Bjornson2015b,massivemimobook}), but not under correlated BS distortion (to the best of our knowledge).}
 To apply DA-MMSE, it is sufficient for the BS to know the channel $\vect{D} \vect{h}_k$ and the correlation matrix $\vect{C}_{z z}+  \sigma^2 \vect{I}_M$ of the received signal in \eqref{eq:distorted-received-signal}.

\begin{observation}
The  BS distortion correlation affects the SINR and can be utilized in the receive combining. The direction of the SE-maximizing receive combining is changed by the distortion correlation.
\end{observation}

Substituting \eqref{eq:combining-vector} into \eqref{eq:SINR} yields
\begin{align} 
\!\!\!\!\gamma_k= p \vect{h}_k^{\Htran} \vect{D}^{\Htran}  \bigg( \!\sum_{i=1,i \neq k}^K \!\!\!p \vect{D}\vect{h}_i \vect{h}_i^{\Htran} \vect{D}^{\Htran} \!+\! \vect{C}_{\eta \eta} \! +\!  \sigma^2 \vect{I}_M \bigg)^{\!-1}  \!\!\!\! \vect{D} \vect{h}_k.\!\!
\end{align}
Recall that $\vect{H}$ is just one channel realization from some arbitrary ergodic fading process. Hence, the ergodic achievable SE $\mathcal{I}(s_k; \vect{v}_k^{\Htran}\vect{y}, \vect{H}) = \mathbb{E}_{\vect{H}}\{\mathcal{I}(s_k; \vect{v}_k^{\Htran}\vect{y}) \} $ over the fading channel in \eqref{eq:distorted-received-signal} is lower bounded by
\begin{equation}  
\mathbb{E}_{\vect{H}}\{\mathcal{I}(s_k; \vect{v}_k^{\Htran}\vect{y}) \} \ge \mathbb{E}_{\vect{H}} \{ \log_2 ( 1+ \gamma_k) \} \label{SE-ideal-tx}
\end{equation}
where $\mathbb{E}_{\vect{H}}\{ \cdot \}$ denotes the expectation with respect to $\vect{H}$.

\subsection{Spectral Efficiency with UE Hardware Impairments}
\label{subsec:SE-non-ideal-tx}

In practice, there are hardware impairments at both the BS and UEs. To quantify the relative impact of both impairments, we next assume that $s_k = \varsigma_k + \omega_k$ for $k=1,\ldots,K$, where $\varsigma_k \sim \CN(0, \kappa p)$ is the actual desired signal from UE $k$ and $\omega_k \sim \CN(0, (1-\kappa) p)$ is a distortion term. The parameter $\kappa \in [0,1]$ determines the level of hardware impairments at the UE side, potentially after some predistortion algorithm has been applied. For analytical tractability, we assume that $\varsigma_k$ and $\omega_k$ are independent, thus the transmit power is $\mathbb{E}\{ |s_k|^2\} = \kappa p + (1-\kappa) p = p$ irrespective of $\kappa$. The independence can be viewed as a worst-case assumption \cite{Hassibi2003a,massivemimobook}, but is mainly made to 
apply the same methodology as above to obtain the achievable SE
\begin{equation}
\mathcal{I}(\varsigma_k; \vect{v}_k^{\Htran}\vect{y}, \vect{H}) = \mathbb{E}_{\vect{H}}\{\mathcal{I}(\varsigma_k; \vect{v}_k^{\Htran}\vect{y}) \} \geq \mathbb{E}_{\vect{H}}\{ \log_2(1+\gamma_k^\prime)\}
\end{equation}
with
\begin{align} \notag 
&\gamma_k^\prime= \\ &
\frac{ \kappa p \vect{v}_k^{\Htran} \vect{D}\vect{h}_k \vect{h}_k^{\Htran} \vect{D}^{\Htran}  \vect{v}_k}{  \vect{v}_k^{\Htran} \big( \! \sum\limits_{i \neq k} p \vect{D}\vect{h}_i \vect{h}_i^{\Htran} \vect{D}^{\Htran} \!+ (1\!-\!\kappa) p\vect{D}\vect{h}_k \vect{h}_k^{\Htran} \vect{D}^{\Htran}  \! + \vect{C}_{\eta \eta}  +  \sigma^2 \vect{I}_M \big)  \vect{v}_k}. \label{SE-non-ideal-tx}
\end{align}
This SINR is also maximized by DA-MMSE in \eqref{eq:combining-vector}, as it can be proved using \cite[Lemma B.4]{massivemimobook}. The reason is that the desired signal and UE distortion are received over the same channel $\vect{D}\vect{h}_k$, thus such distortion cannot be canceled by receive combining without also canceling the desired signal. 

\begin{observation}
The UE distortion does not change the SE-maximizing receive combining vector at the BS.
\end{observation}

\section{Quantifying the Impact of Non-linearities}
\label{sec:non-linearities}

\begin{figure}
  \centering  
    \includegraphics[width=0.5\textwidth]{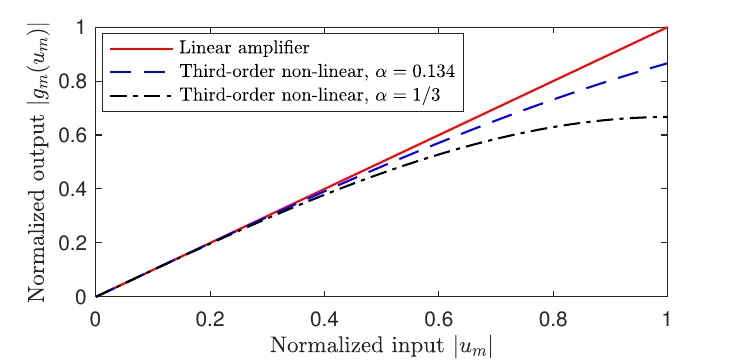} 
      \caption{Comparison between a linear amplifier and the non-linear amplifiers described by \eqref{eq:third-order-nonlinear} with $\boff=1$.} \label{figure_amplifier}  
\end{figure}

The BS distortion term $\vect{v}_k^{\Htran} \vect{C}_{\eta \eta}   \vect{v}_k$ appears in \eqref{eq:SINR} and \eqref{SE-non-ideal-tx}.
To analyze the characteristics of this term and, particularly, the impact of the distortion correlation (i.e., the off-diagonal elements in $\vect{C}_{\eta \eta}$), we now focus on the AM-AM distortion caused by the LNA. This can be modeled, in the complex baseband, by a third-order memoryless non-linear function \cite{Schenk2008a}
\begin{equation} \label{eq:third-order-nonlinear}
g_m(u_m) = u_m - a_m |u_m|^2 u_m, \quad  m=1,\ldots,M.
\end{equation} 
This is a valid model of amplifier saturation when $a_m \geq 0$ and for such input amplitudes $|u_m|$ that $|g_m(u_m)|$ is an increasing function.
This occurs for $|u_m| \leq \frac{1}{\sqrt{3a_m}}$, while clipping occurs for input signals with larger amplitudes. In practice, the LNA is operated with a backoff to avoid clipping and limit the impact of the non-linear amplification.
The value of $a_m$ depends on the circuit technology and how we normalize the output power of the LNA. We can model it as \cite{3GPP_PA_models}
\begin{equation} \label{eq:a_m_model}
a_m = \frac{\alpha}{\boff\mathbb{E}\{ |u_m|^2 \} }
\end{equation}
where $\mathbb{E}\{ |u_m|^2 \} $ is the average signal power and $\boff \geq 1$ is the back-off parameter selected based on the peak-to-average-power ratio (PAPR) of the input signal to limit the risk for clipping. The parameter $\alpha>0$ determines the non-linearities for normalized input signals with amplitudes in $[0,1]$. The worst case is given by $\alpha= 1/3$, for which the LNA saturates at unit input amplitude. A smaller value of $\alpha = 0.1340$ was reported in \cite{3GPP_PA_models} for a GaN amplifier operating at 2.1\,GHz. These amplifiers are illustrated in Fig.~\ref{figure_amplifier} for $\boff=1$. 

We can use this model to compute $\vect{C}_{\eta \eta}$ in closed form. We let $\rho_{ij} = \EcondH\{ u_i u_j^*\} = [\vect{C}_{uu}]_{ij}$ denote the $(i,j)$th element of $\vect{C}_{uu}$.
With this notation, we have
\begin{equation}
d_m = \frac{\EcondH\{ g_m(u_m) u_m^\star \}}{  \EcondH \{ | u_m |^2 \}} = \frac{\rho_{mm} - 2 a_m \rho_{mm}^2}{\rho_{mm}}= 1 - 2 a_m \rho_{mm}
\end{equation}
for $m=1,\ldots,M$ and thus
\begin{equation} \label{eq:DCssD}
[\vect{D} \vect{C}_{uu} \vect{D}^{\Htran}]_{ij} = d_i \rho_{ij} d_j^* = (1 - 2 a_i \rho_{ii}) \rho_{ij} (1 - 2 a_j \rho_{jj}).
\end{equation}

\begin{lemma} \label{lemma:nonlinear_distortion_matrix}
With the third-order non-linearities in \eqref{eq:third-order-nonlinear}, the distortion term's correlation matrix in \eqref{eq:Cetaeta} is given by
\begin{align} \label{eq:Cetaeta_expression1}
 [\vect{C}_{\eta \eta} ]_{ij} = 2 a_i a_j  | \rho_{ij}|^2  \rho_{ij}.
\end{align}
\end{lemma}
\begin{IEEEproof}
The proof is given in Appendix~\ref{app:lemma:nonlinear_distortion_matrix}.
\end{IEEEproof}

The BS distortion correlation matrix in Lemma~\ref{lemma:nonlinear_distortion_matrix} can be expressed in matrix form as 
\begin{align} \label{eq:Cetaeta_expression}
\vect{C}_{\eta \eta} = 2 \vect{A} \left( \vect{C}_{uu} \odot \vect{C}_{uu}^* \odot \vect{C}_{uu}  \right) \vect{A}
\end{align}
where $\vect{A} = \diag(a_1,\ldots,a_M)$.
Since the information signal $\vect{u}$ typically has correlated elements (i.e., $ \vect{C}_{uu}$ has non-zero off-diagonal elements),  \eqref{eq:Cetaeta_expression} implies that also the distortion has correlated elements. The correlation coefficient between $u_i$ and $u_j$ is
\begin{equation}
\xi_{u_i u_j} = \frac{\EcondH\{ u_i u_j^*\} }{\sqrt{\EcondH\{ |u_i|^2\}  \EcondH\{  |u_j|^2\} }} = \frac{\rho_{ij}}{\sqrt{\rho_{ii} \rho_{jj}}}
\end{equation}
while the correlation coefficient between $\eta_i$ and $\eta_j$ is
\begin{equation}
\xi_{\eta_i \eta_j} =\! \frac{\EcondH\{ \eta_i \eta_j^*\} }{\sqrt{\EcondH\{ |\eta_i|^2\}  \EcondH\{  |\eta_j|^2\} }} = \! \frac{|\rho_{ij}|^2 \rho_{ij}}{\sqrt{\rho_{ii}^3 \rho_{jj}^3}} = |\xi_{u_i u_j} |^2 \xi_{u_i u_j}.
\end{equation}
Clearly, $|\xi_{\eta_i \eta_j} | = |\xi_{u_i u_j} |^3 \leq |\xi_{u_i u_j} |$ since $ |\xi_{u_i u_j} | \in [0,1]$.

\begin{observation} \label{observ:less_correlated_than_signal}
The distortion terms are less correlated than the corresponding signal terms.
\end{observation}

While this observation applies to the uplink, similar results for the downlink can be found in \cite{Moghadam2012a,Mollen2018b}.

\subsection{Quantifying the Distortion Terms With Non-Linearities}

If the BS distortion terms are only weakly correlated, it would be analytically tractable to neglect the correlation. This effectively means using the diagonal correlation matrix
\begin{equation} \label{Cee_diag}
\vect{C}_{\eta \eta}^{\diag} = \vect{C}_{\eta \eta}  \odot \vect{I}_M
\end{equation}
which has the same diagonal elements as $ \vect{C}_{\eta \eta}$. This simplification is made in numerous papers that analyze SE \cite{Bai2013a,Orhan2015a,Bjornson2014a,massivemimobook}. We will now quantify the impact that such a simplification has when the BS distortion is caused by the third-order non-linearity in \eqref{eq:third-order-nonlinear}. 
For this purpose, we consider i.i.d.~Rayleigh fading channels $\vect{h}_k \sim \CN (\vect{0},\vect{I}_M)$ for $k=1,\ldots,K$. The average power received at BS antenna $m$ in \eqref{eq:a_m_model} is
\begin{equation}
\mathbb{E}\{ |u_m|^2 \} = \mathbb{E} \left\{ p \sum_{k=1}^{K} | h_{km} |^2 \right\} = p K.
\end{equation}
The impact of distortion correlation can be quantified by considering the distortion term $\vect{v}_k^{\Htran}\vect{C}_{\eta \eta} \vect{v}_k$   in \eqref{eq:SINR} and \eqref{SE-non-ideal-tx} and comparing it with $\vect{v}_k^{\Htran}\vect{C}_{\eta \eta}^{\diag} \vect{v}_k$ where the correlation is neglected. To make a fair comparison, we consider MR combining with
\begin{equation} \label{eq:MR-combining}
\vect{v}_k = \frac{\vect{h}_k}{ \sqrt{\mathbb{E}\{ \| \vect{h}_k  \|^2\}}}
\end{equation}
which does not suppress distortion
(in Section~\ref{subsec:distortion-directivtity} we showed that BS distortion correlation can be exploited to reject distortion by receive combining, but that is not utilized here).

\begin{lemma} \label{lemma:ratios-of-distortion}
Consider i.i.d.~Rayleigh fading channels and $a_m$ given by  \eqref{eq:a_m_model},
then
\begin{align} \label{eq:ratio-of-distortion}
&\frac{\mathbb{E}\{ \vect{h}_k^{\Htran} \vect{C}_{\eta \eta} \vect{h}_k\}}{\mathbb{E}\{ \| \vect{h}_k \|^2 \} } = \frac{2 \alpha^2 p}{\boff
^2} \! \left( K+6+\frac{9}{K}+\frac{4}{K^2}+\frac{2M(K+1)}{K^2} \right)\!
\end{align}
which is approximated by
\begin{align} \label{eq:ratio-of-distortion-uncorr}
 \frac{\mathbb{E}\{ \vect{h}_k^{\Htran} \vect{C}_{\eta \eta}^{\diag} \vect{h}_k\}}{\mathbb{E}\{ \| \vect{h}_k \|^2 \} } = \frac{2 \alpha^2 p}{\boff^2} \! \left( K+6+\frac{11}{K}+\frac{6}{K^2} \right)
\end{align}
if the BS distortion correlation is neglected.
\end{lemma}
\begin{IEEEproof}
The proof is given in Appendix~\ref{app:lemma:ratios-of-distortion}.
\end{IEEEproof}

The BS distortion term in \eqref{eq:ratio-of-distortion-uncorr} without correlation is independent of the number of antennas, which implies that the distorted signal components are non-coherently combined by MR combining. In contrast, the correlated BS distortion term in \eqref{eq:ratio-of-distortion} contains an additional term $2(M-1)(K+1)/K^2$ that grows linearly with the number of antennas.  Hence, the correlation creates extra distortion that is also coherently combined by MR combining, as also observed in \cite{Mollen2018a}.

In both cases, there is one component that grows with $K$ and several components that reduces with $K$. Hence, we can expect the distortion terms to be unimodal functions of $K$, such that they are first reducing and then increasing with $K$.

The average distortion power with MR combining is larger when the distortion terms are correlated, since the fraction
\begin{equation} \label{eq:distortion-fraction}
\!\!\!\frac{ \frac{\mathbb{E}\{ \vect{h}_k^{\Htran} \vect{C}_{\eta \eta} \vect{h}_k\}}{\mathbb{E}\{ \| \vect{h}_k \|^2 \} } }{ \frac{\mathbb{E}\{ \vect{h}_k^{\Htran} \vect{C}_{\eta \eta}^{\diag} \vect{h}_k\}}{\mathbb{E}\{ \| \vect{h}_k \|^2 \} } } = \frac{ \mathbb{E}\{ \vect{h}_k^{\Htran} \vect{C}_{\eta \eta} \vect{h}_k\} }{ \mathbb{E}\{ \vect{h}_k^{\Htran} \vect{C}_{\eta \eta}^{\diag} \vect{h}_k\} } 
= 1 + \frac{ 2(M-1) }{(K+2)(K+3)}\!\!
\end{equation}
is larger than one and also independent of $\alpha$ and $\boff$. The size of the second term depends on the relation between $M$ and $K$; it is linearly increasing with $M$ and quadratically decreasing with $K$.

In addition to the BS distortion term, the denominator of the SINR in \eqref{SE-non-ideal-tx} also contains the term $(1-\kappa) p | \vect{h}_k^{\Htran} \vect{D}^{\Htran} \vect{v}_k|^2$ which originates from the UE distortion. When using MR combining in \eqref{eq:MR-combining}, it can be computed as follows.

\begin{lemma} \label{lemma:UE-distortion-term}
Consider i.i.d.~Rayleigh fading channels and $a_m$ given by  \eqref{eq:a_m_model},
then
\begin{align} \notag
&\frac{\mathbb{E}\{ | \vect{h}_k^{\Htran} \vect{D} \vect{h}_k |^2\}}{\mathbb{E}\{ \| \vect{h}_k \|^2 \} } = 
(M+1)-\frac{4\alpha (MK+K+M+3)}{\boff K} \\ &+\frac{4\alpha^2 (MK^2+8K+11+2MK+K^2+M)}{\boff^2 K^2}. \label{eq:UE-distortion-term}
\end{align}
\end{lemma}
\begin{IEEEproof}
The proof is given in Appendix~\ref{app:lemma:UE-distortion-term}.
\end{IEEEproof}

The first term in \eqref{eq:UE-distortion-term} is dominant since $\alpha/\boff< 1$. Hence, the UE distortion will basically grow linearly with $M$, similar to the correlated BS distortion term in \eqref{eq:ratio-of-distortion}.
The UE distortion is also affected by $K$, but the impact is rather small since $K$ only appears in the two non-dominant terms. We will show numerically that the UE distortion grows with $K$.

\subsection{What Happens if the Distortion Correlation is Neglected?}
\label{subsec:distortion-correlation-neglected}

\begin{figure}
  \centering 
    \includegraphics[width=0.5\textwidth]{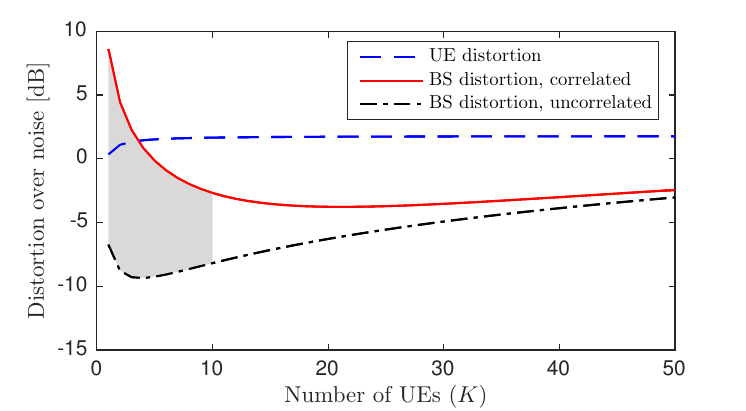} 
      \caption{The BS and UE distortion power over noise with and without BS distortion correlation. The approximation error drops significantly in the shaded interval. The UE distortion dominates for $K\geq 5$ in this setup with $M=200$.} \label{figure_distortioncomparison} 
\end{figure}

We will now quantify the size of the BS and UE distortion terms.  We consider a Massive MIMO setup with $M=200$, a worst-case LNA with $\alpha = 1/3$, $\boff = 7$\,dB, and an SNR of $p/\sigma^2=0$\,dB. We consider high-quality transmitter hardware with $\kappa=0.99$ \cite[Sect.~6.1.2]{massivemimobook} and the signal-to-distortion power ratio $\kappa/(1-\kappa)=99$, which is higher than $[\vect{D} \vect{C}_{uu} \vect{D}^{\Htran}]_{ii} /  [\vect{C}_{\eta \eta} ]_{ii} \approx 85$  for the LNA. The solid and dash-dotted curves in Fig.~\ref{figure_distortioncomparison} show \eqref{eq:ratio-of-distortion} and \eqref{eq:ratio-of-distortion-uncorr}, normalized by the noise, as a function of $K$. The dashed curve in Fig.~\ref{figure_distortioncomparison} shows the UE distortion based on \eqref{eq:UE-distortion-term}.

The BS distortion terms are first reducing with $K$ and then slowly increasing again, since the LNA distortion originates from all the UEs and their total transmit power is $pK$. The correlation has a huge impact on the BS distortion term when there are few UEs. 
Quantitively speaking, it is more than 10\,dB larger than without correlation. The reason is that the correlation gives the distortion vector $\boldsymbol{\eta}$ a similar direction as $\vect{h}_k$, for $k=1,\ldots,K$, when $K$ is small. Hence, the distortion effect is amplified by MR combining. The gap to the curve with uncorrelated distortion reduces with $K$. In the shaded part, the gap reduces from $15.3$ to $5.5$\,dB.
This is expected from \eqref{eq:distortion-fraction}, since the elements of $\boldsymbol{\eta}$ becomes less correlated when $K$ grows. 

The UE distortion is a slowly increasing function of $K$ in Fig.~\ref{figure_distortioncomparison}. The correlated BS distortion is the dominant factor for $K\leq 3$, but for larger values of $K$ (as is the case in Massive MIMO), the UE distortion becomes much higher (5\,dB larger in this example). The reason is that many of the term in the BS distortion expression (particularly the ones that depend on $M$) reduce with $K$.

\begin{observation}
The correlation of the BS distortion reduces with $K$. The BS distortion will eventually have a smaller impact than the UE distortion, which doesn't reduce with $K$.
\end{observation}

The first part of this observation is in line with the downlink analysis in \cite{Moghadam2012a,Mollen2018b,Larsson2018a} and the uplink analysis in \cite{Mollen2018a}, while the second part is a new observation. Note that the prior works did not quantify the impact of BS distortion on the SE.

\subsection{Distortion Directivity with One or Multiple UEs}
\label{subsec:distortion-directivtity}

When there is only one UE in the network, the $M \times M$ signal correlation matrix $\vect{D}\vect{C}_{uu} \vect{D}^{\Htran}= p \vect{D}\vect{h}_1 \vect{h}_1^{\Htran} \vect{D}^{\Htran}$ has rank one. This property carries over to the distortion term's correlation matrix in \eqref{eq:Cetaeta_expression}, which becomes
\begin{equation}
\vect{C}_{\eta \eta} = \frac{2\alpha^2 p}{\boff^2} \tilde{\vect{h}}_1 \tilde{\vect{h}}_1^{\Htran}
\end{equation}
where $\tilde{\vect{h}}_1 = [ |h_{11}|^2 h_{11} \, \ldots \, | h_{1M} |^2 h_{1M}]^{\Ttran} \in \mathbb{C}^{M}$ is the eigenvector corresponding to the only non-zero eigenvalue. 
Due to the AM-AM distortion, the $m$th element of $\vect{D}\vect{h}_1$ and $\tilde{\vect{h}}_1$  have the same phase, but different amplitudes. In the special case of far-field free-space propagation, which is sometimes misleadingly referred to as line-of-sight (LoS) propagation, the elements of $\vect{h}_1$ have equal amplitude and, hence, $\vect{D}\vect{h}_1$ and $\tilde{\vect{h}}_1$ become parallel. However, in practice, the multi-path propagation leads to substantial amplitude variations in both LoS and non-LoS scenarios, as demonstrated by the measurement results in \cite{Gao2015b}. Hence, $\vect{D}\vect{h}_1$ and $\tilde{\vect{h}}_1$ are generally non-parallel vectors when $M\geq 2$. 

\begin{observation}
For $K=1$ and $M\geq 2$, the BS distortion vector $\boldsymbol{\eta}$ has a related but different direction than the desired signal vector $\vect{D}\vect{h}_1 s_1$.
\end{observation}

This has important implications on the receive combining. As a baseline scheme, we use 
\begin{equation} \label{eq:DA-MR}
\vect{v}_k = \frac{ \vect{D} \vect{h}_k}{\| \vect{D} \vect{h}_k\|}
\end{equation}
for UE~$k$, since the effective channel in \eqref{eq:distorted-received-signal} is $\vect{D} \vect{h}_k$. We call this \emph{distortion-aware MR (DA-MR) combining} to differentiate it from conventional MR in \eqref{eq:MR-combining} which does not take $\vect{D}$ into account. With this scheme, the BS distortion term $\vect{v}_1^{\Htran} \vect{C}_{\eta \eta} \vect{v}_1$ grows linearly with $M$, as it can be inferred from \eqref{eq:ratio-of-distortion} or  \cite{Mollen2018a}.

However, we can instead take the channel vector $\vect{D}\vect{h}_1$ and project it onto the orthogonal subspace of $\tilde{\vect{h}}_1$ to obtain the receive combining vector
\begin{equation}
\vect{v}_1 = \frac{\left( \vect{I}_M - \frac{1}{\| \tilde{\vect{h}}_1\|^2} \tilde{\vect{h}}_1 \tilde{\vect{h}}_1^{\Htran}  \right) \vect{D}\vect{h}_1}{\left\| \left( \vect{I}_M - \frac{1}{\| \tilde{\vect{h}}_1\|^2} \tilde{\vect{h}}_1 \tilde{\vect{h}}_1^{\Htran}  \right) \vect{D}\vect{h}_1  \right\|}.
\end{equation}
For $M\geq 2$, this results in $\vect{v}_1^{\Htran} \vect{C}_{\eta \eta} \vect{v}_1=0$ and 
$\vect{v}_1^{\Htran} \vect{D}\vect{h}_1 \neq 0$. We call this approach \emph{distortion-aware zero-forcing (DA-ZF) combining}. It can be generalized to a multiuser case by taking the channel of a given UE and projecting it orthogonally to the subspace jointly spanned by the co-user channels and $\vect{C}_{\eta \eta}$.
 
\begin{figure}
  \centering 
    \includegraphics[width=0.5\textwidth]{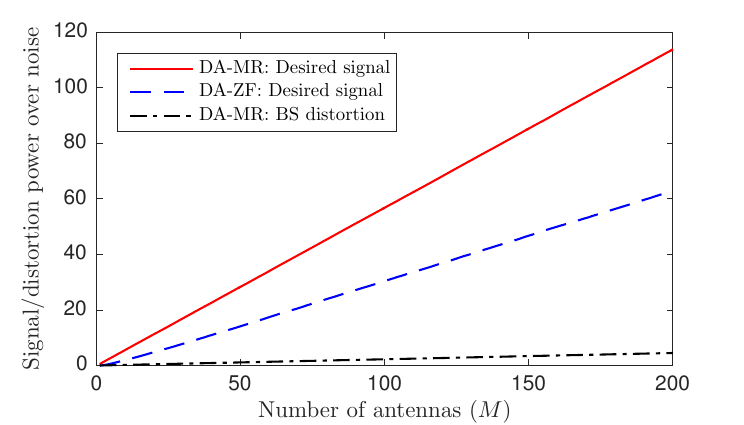} 
      \caption{The desired signal and BS distortion (due to non-linearities) grow linearly with $M$ when using DA-MR, while DA-ZF can cancel the BS distortion while keeping a linearly increasing desired signal.} \label{figure_distortion_singleuser}  
\end{figure}

Fig.~\ref{figure_distortion_singleuser} compares the desired signal term $\kappa p |\vect{v}_1^{\Htran} \vect{D}\vect{h}_1 |^2$ and BS distortion term $\vect{v}_1^{\Htran} \vect{C}_{\eta \eta} \vect{v}_1$ (normalized by the noise power) with DA-MR and the corresponding desired signal term achieved by DA-ZF. These terms are shown as a function of $M$ for $K=1$, i.i.d.~Rayleigh fading, $\alpha = 1/3$, $\boff = 7$\,dB, $\kappa=0.99$, and SNR $p/\sigma^2=0$\,dB. 

With DA-MR, both the signal and distortion grow linearly with $M$, which will lead to SINR saturation in \eqref{SE-non-ideal-tx} as $M\to \infty$, as previously observed in \cite{Mollen2018a}. In contrast, when using DA-ZF, the BS distortion is forced to zero in the SINR expression, while the desired signal is still growing linearly with $M$. Hence, DA-ZF removes the SINR saturation due to BS distortion. The price to pay is a loss in signal power, which is approximately 50\% smaller than with DA-MR.

\begin{observation}
When the BS distortion is correlated, it can be suppressed by receive combining by sacrificing a part of the array gain.
\end{observation}

The DA-ZF scheme was only introduced to demonstrate this key property, but is not needed in practice. DA-MMSE will find the SE-maximizing tradeoff between achieving a strong signal power and suppressing interference and distortion.

\begin{figure}
  \centering 
    \includegraphics[width=0.5\textwidth]{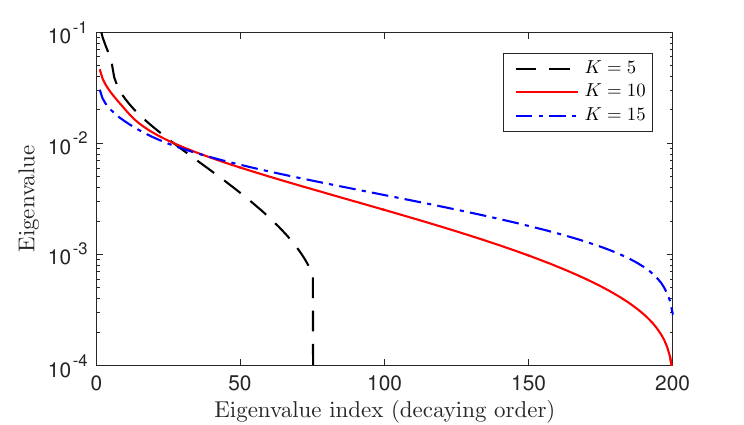} 
      \caption{Ordered eigenvalues of the BS distortion correlation matrix $\vect{C}_{\eta \eta}$ (due to non-linearities) for $M=200$ and varying $K$. The rank of $\vect{C}_{\eta \eta}$ increases rapidly with $K$ and the difference between the largest and smallest non-zero eigenvalues also reduces when $K$ grows.} \label{figure_eigenvalues}  
\end{figure}

When $K$ increases, the ranks of $\vect{D}\vect{C}_{uu} \vect{D}^{\Htran}$ and  $\vect{C}_{\eta \eta}$ also grow. While the rank of the signal correlation matrix equals $K$, the rank of the distortion correlation matrix grows substantially faster, as one might expect from Observation~\ref{observ:less_correlated_than_signal}. This is illustrated in Fig.~\ref{figure_eigenvalues} for the same setup as in the last figure, but with $M=200$ and $K \in \{5, \, 10, \, 15 \}$. This figure shows the eigenvalues of $\vect{C}_{\eta \eta}/\tr(\vect{C}_{\eta \eta})$ in decaying order, averaged over different i.i.d.~Rayleigh fading realizations. There are 75 non-zero eigenvalues when $K=5$, while all the 200 eigenvalues are non-zero when $K=10$. As $K$ continues to increase, the difference between the largest and smallest eigenvalues reduces.
This uplink result is in line with previous downlink results in \cite{Mollen2018b} that quantify how many UEs are needed to approximate $\vect{C}_{\eta \eta}$ as a scaled identity matrix.

\section{Quantifying the Impact of Quantization}
\label{sec:quantization}

The finite-resolution quantization in the ADCs is another major source of hardware distortion at the receiving BS. The real and imaginary parts are quantized independently by a quantization function $Q(\cdot)$ using $b$ bits. The $L=2^b$ quantization levels, $\ell_1,\ldots,\ell_L$, are symmetric around the origin such that $\ell_n = -\ell_{L-n+1}$ for $n=1,\ldots,L$. The thresholds are denoted $\tau_0,\ldots,\tau_L$ (with $\tau_0=-\infty$, $\tau_L=\infty$), such that
\begin{equation}
Q(x) = \ell_n \quad \textrm{if} \,\,\, x\in [\tau_{n-1}, \tau_{n}), \quad n=1,\ldots,L.
\end{equation}

Using this general quantization model, the matrices $\vect{D}$ and  $\vect{C}_{\eta \eta}$ from Section~\ref{sec:system-model} can be characterized. The elements $d_1,\ldots,d_M$ of $\vect{D}$ follow directly from \cite[Th.~1]{Jacobsson2017b} and become
\begin{align} \label{D_matrix_quant}
d_m &= \sum_{n=1}^{L} \frac{\ell_n}{\sqrt{\pi \rho_{mm}}} \left( e^{-\frac{\tau_{n-1}^2}{\rho_{mm}}} - e^{-\frac{\tau_{n}^2}{\rho_{mm}}}  \right).
\end{align}
The diagonal elements of $\vect{C}_{zz}$ are computed as
\begin{align} \notag
&[\vect{C}_{zz}]_{ii} \\ \notag &
= \mathbb{E}\left\{ \left( Q(\Re(u_i)) + \imagunit Q(\Im(u_i))  \right) \left( Q(\Re(u_i)) - \imagunit Q(\Im(u_i))  \right) \right\} \\ \notag 
&\overset{(a)}{=} 2 \mathbb{E}\left\{  Q(\Re(u_i)) Q(\Re(u_i))   \right\} =  \int_{-\infty}^{\infty} \frac{2 \big( Q(z) \big)^2 }{\sqrt{\pi \rho_{ii}}} e^{\frac{-z^2}{\rho_{ii}}} dz \\ \notag
&= \sum_{n=1}^{L} \frac{2 \ell_n^2}{\sqrt{\pi \rho_{ii}}} \int_{\tau_{n-1}}^{\tau_n} e^{\frac{-z^2}{\rho_{ii}}} dz\\ \label{Czz_derivation2} 
&= \sum_{n=1}^{L}  \ell_n^2 \left( \erf\left( \frac{\tau_n}{\sqrt{\rho_{ii}}}  \right) - \erf\left( \frac{\tau_{n-1}}{\sqrt{\rho_{ii}}}  \right)  \right) 
\end{align}
where $(a)$ follows from the independence of the real and imaginary part of $u_i$, that $\mathbb{E}\{ Q(\Re(u_i))\} = \mathbb{E}\{ Q(\Im(u_i))\} = 0$ due to the quantization level symmetry, and by exploiting that $\Re(u_i) \sim \mathcal{N}(0,\rho_{ii}/2)$. The remaining steps follow from direct computation.

The off-diagonal elements of $\vect{C}_{zz}$ can be computed in the same way, but do not lead to closed-form expressions (the expectation of error functions of random variables lacks an analytical solution). This is, at least, an indication of the existence of distortion correlation, since otherwise the off-diagonal elements would simply match the corresponding elements in $\vect{D} \vect{C}_{uu} \vect{D}^{\Htran}$.
In what follows, we will compute $\vect{C}_{zz}$ by Monte-Carlo methods to quantify the distortion correlation.

\subsection{Correlation in Quantization Distortion}

To quantify the impact of distortion correlation in the quantization, we consider  i.i.d.~Rayleigh fading channels with $M=100$ antennas and either $K=1$ or $K=2$ UEs. ADC resolutions $b \in \{1,2,\ldots,8\}$ are considered and 
the quantization thresholds are optimized numerically using the Lloyd algorithm \cite{Lloyd1982a} for each $b$. 

\begin{figure}
  \centering 
    \includegraphics[width=0.5\textwidth]{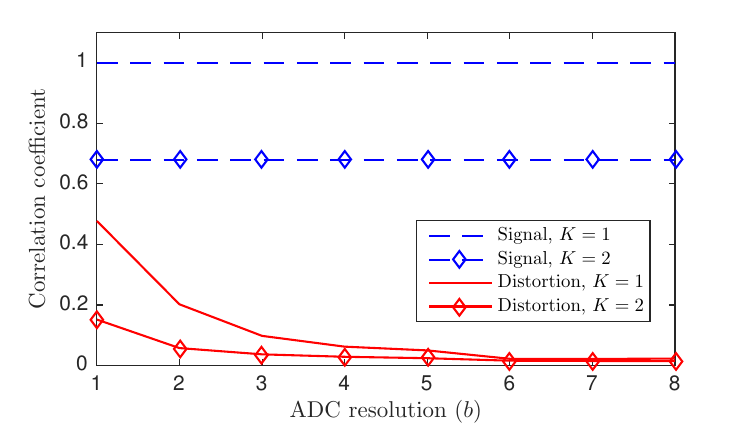} 
      \caption{Average correlation coefficients
$\xi_{u_i u_j}$ for the signal and $\xi_{\eta_i \eta_j}$ for the quantization distortion as a function of the ADC resolution $b$.} \label{figure_quantization_correlation}  
\end{figure}

Fig.~\ref{figure_quantization_correlation} shows the correlation coefficients
$\xi_{u_i u_j}$ for the signal and $\xi_{\eta_i \eta_j}$ for the distortion, where we have averaged over different channel realizations. For $K=1$, we have $\xi_{u_i u_j}=1$, but since the Rayleigh fading channel gives different phases to $u_i$ and $u_j$, their real and imaginary parts are different; thus, the quantization distortion correlation is substantially smaller. While the signal correlation is independent of the ADC resolution,  the distortion correlation decays rapidly when $b$ increases. The distortion correlation is nearly zero for $b\geq 6$ with $K=1$ and for $b\geq 4$ with $K=2$. If we would continue increasing $K$, the correlation between $u_i$ and $u_j$ will decrease, which also leads to less correlation between the distortion terms $\eta_i$ and $\eta_j$. This is similar to the decorrelation with the number of UEs that we observed for non-linearities in Section~\ref{sec:non-linearities}.

\begin{figure}
  \centering 
    \includegraphics[width=0.5\textwidth]{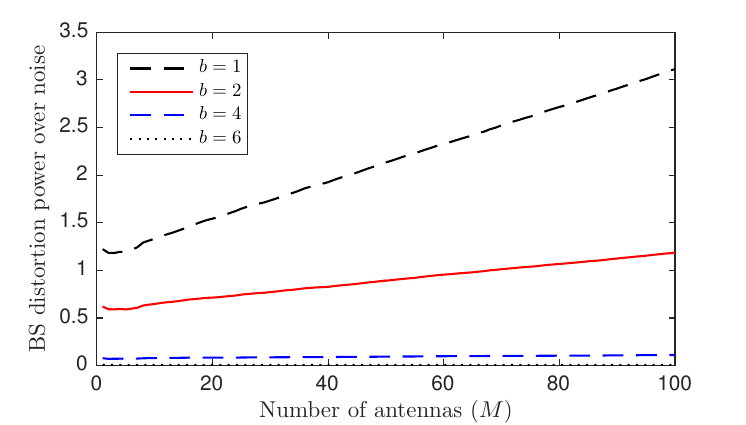} 
      \caption{The BS distortion power (due to quantization) over noise for $K=5$. 
The distortion power grows linearly with $M$ when using MR combining, due to distortion correlation, but the slope and absolute power decay rapidly with the ADC resolution $b$.} \label{figure_quantization_correlation_M}  
\end{figure}

The distortion correlation leads to eigenvalue variations in  $\vect{C}_{\eta \eta} $. 
The correlation due to quantization has similar impact as the correlation due to non-linearities: the dominating eigenvectors are similar to the channel vectors.
To demonstrate this, Fig.~\ref{figure_quantization_correlation_M} shows $\frac{\mathbb{E}\{ \vect{h}_k^{\Htran} \vect{C}_{\eta \eta} \vect{h}_k\}}{\mathbb{E}\{ \| \vect{h}_k \|^2 \} }$ (normalized by the SNR $p/\sigma^2=0$\,dB) as a function of the number of antennas. This represents the power of the BS distortion term in the SINR when using MR combining. We consider $K=5$, i.i.d.~Rayleigh fading, and a varying number of quantization bits. In all cases, the distortion term grows linearly with $M$, which means that the quantization errors at the different antennas are (partially) coherently combined---otherwise, the curves would be flat. However, both the slope and the distortion power decay rapidly as $b$ increases. For $b\geq 4$, the linear increase is barely visible, and for $b\geq 6$, the distortion power is negligible as compared to the noise.

At what point the quantization distortion correlation can be neglected in the SINR computation depends on the SNRs of the UEs. If one of the $K$ UEs has a much higher SNR than the others, then the impact of distortion correlation becomes similar to the $K=1$ case. Hence, based on the previous figures, we make the following observation that applies for any $K$.

\begin{observation}
The quantization distortion is correlated between antennas, but the impact of the correlation is negligible for bit resolutions $b\geq 6$ and the entire quantization distortion term is negligible at typical SNRs.
\end{observation}

Since the energy consumption of a $6$\,bit ADC with a sampling frequency of a few GHz is only a few mW \cite{Chan2015a,Choo2016a}, one can use a hundred of them in Massive MIMO without an appreciable impact on the total energy consumption. In other words, by using 6\,bit ADCs, one can jointly achieve a negligible quantization distortion and energy consumption, so there is no need to consider lower ADC resolutions than that.

The quantization distortion correlation is often neglected in the literature (cf.~\cite{Bai2013a,Orhan2015a}) and this is a valid approximation when considering high/medium-resolution ADCs. However, since the Massive MIMO literature contains many papers on low-resolution ADCs (cf.~\cite{Jacobsson2017b,Studer2016a,Mollen2017a,Verenzuela2017b}), it is not necessarily a good approximation in every situation. In the next section, we will investigate the joint impact of distortion from non-linearities and quantization on the SE.

\section{How Much is the SE Affected by Neglecting Distortion Correlation at the BS?}
\label{sec:SE-analysis}

In the previous sections, we have shown that the BS distortion caused by non-linearities and quantization is correlated between the antennas. At the same time, we noticed that the correlation from non-linearities has a limited impact on the distortion terms in the SINR for $K \geq 5$ (see Fig.~\ref{figure_distortioncomparison}) and, for typical bit resolutions ($b\geq 4$), the quantization distortion also appears to be small. As stated in the title, the main purpose of this paper is to demonstrate that the distortion correlation has a negligible impact on the SE in Massive MIMO scenarios (i.e., $M>100$ and $K\in[10,50]$ UEs). To do so, we will compute the SE expressions derived in Section~\ref{subsec:SE-non-ideal-tx} numerically in different scenarios, with both non-linearities and quantization errors.

\subsection{Impact of the Number of UEs}

We first consider a varying number of UEs. We assume i.i.d.~Rayleigh fading with $M=100$ antennas and the SNRs are $p/\sigma^2=0$\,dB. The non-ideal hardware at the BS and UEs are represented by $\alpha = 1/3$, $\boff=7$\,dB, $b=6$, and $\kappa=0.99$. As discussed in Section~\ref{subsec:distortion-correlation-neglected}, these parameters give (slightly) higher hardware quality at the UEs than at the BS. This assumption is made in an effort to not underestimate the impact of BS distortion.

\begin{figure}
  \centering  
    \includegraphics[width=0.5\textwidth]{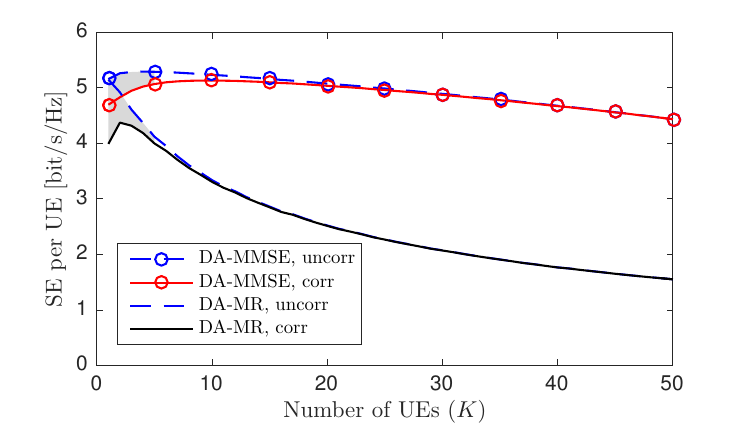} 
      \caption{SE per UE as a function of the number of UEs with $M=100$. We compare DA-MMSE and DA-MR when correlation of the BS distortion is either neglected ($\mathrm{uncorr}$) or accounted for ($\mathrm{corr}$). Every UE has the same SNR in this figure.} \label{figure_SE_users}  
\end{figure}

Fig.~\ref{figure_SE_users} shows the SE per UE, as a function of the number of UEs. We consider either optimal DA-MMSE combining from \eqref{eq:combining-vector} or DA-MR combining from \eqref{eq:DA-MR}. The solid lines in Fig.~\ref{figure_SE_users} represent the exact SE, taking the correlation of the BS distortion into account, while the dashed lines represent approximate SEs achieved by neglecting the distortion correlation; that is, using $\vect{C}_{\eta \eta}^{\diag} $ defined in \eqref{Cee_diag} instead of $\vect{C}_{\eta \eta}$. 
By neglecting the correlation, we get a biased SE that is higher than in practice. However, although the choice of the receive combining scheme has a large impact on the SE, the approximation error is negligible for $K \geq 5$ with both schemes, which is the case in Massive MIMO. 
For $K< 5$, the shaded gap ranges from 9.8\% to 4.3\% when using DA-MMSE, which is still rather small.

\begin{observation}
The distortion correlation has small or negligible impact on the uplink SE in Massive MIMO with i.i.d.~Rayleigh fading and equal SNRs for all UEs.
\end{observation}

This is in line with the Massive MIMO paper \cite{Bjornson2014a} and the book \cite{massivemimobook}, which make similar claims but without providing analytical or numerical evidence.

\subsection{Impact of SNR Differences and Channel Model}

Next, we set $K=5$ and vary the SNRs of the UEs by drawing values from $-10$\,dB to $10$\,dB uniformly at random. The other simulation parameters are kept the same, except that we consider three different channel models:
\begin{enumerate}
\item i.i.d.~Rayleigh fading, as defined above.
\item Correlated Rayleigh fading, where the BS is equipped with a uniform linear array and the UEs are uniformly distributed in the angular interval $[-60^\circ,60^\circ]$ around the boresight. The spatial correlation is computed using the local scattering model from \cite[Sec.~2.6]{massivemimobook} with Gaussian distribution and $10^\circ$ angular standard deviation.
\item Free-space propagation, with the same array type and UE distribution as in 2). This model is also relevant for mmWave communications with one strong signal path.
\end{enumerate}

\begin{figure} 
        \centering
        \begin{subfigure}[b]{\columnwidth} \centering 
                \includegraphics[width=\textwidth]{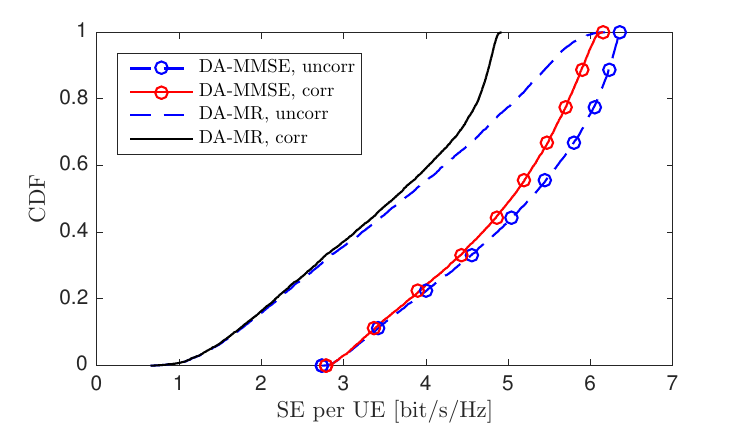} 
                \caption{i.i.d.~Rayleigh fading} 
                \label{figure_SE_pathloss_iid}
        \end{subfigure} 
        \begin{subfigure}[b]{\columnwidth} \centering 
                \includegraphics[width=\textwidth]{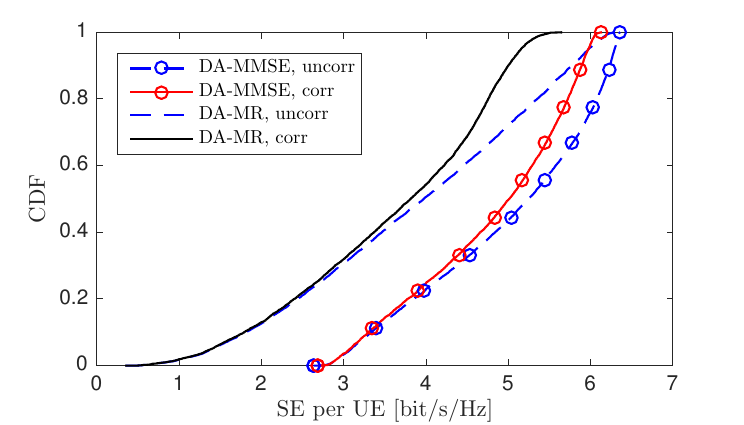} 
                \caption{Correlated Rayleigh fading} 
                \label{figure_SE_pathloss_corr}
        \end{subfigure} 
        \begin{subfigure}[b]{\columnwidth} \centering
                \includegraphics[width=\textwidth]{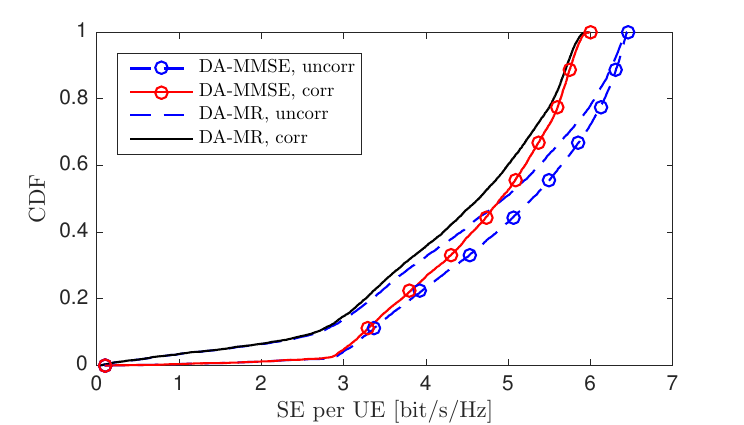} 
                \caption{Free-space propagation}  
                \label{figure_SE_pathloss_fsp} 
        \end{subfigure} 
        \caption{CDF of the SE per UE with DA-MMSE and DA-MR when the correlation of the BS distortion is either neglected ($\mathrm{uncorr}$) or accounted for ($\mathrm{corr}$). The randomness is due to the SNRs being  uniformly distributed from $-10$\,dB to $10$\,dB. Three different channel models are considered with $K=5$ and $M=100$.}
        \label{figure_SE_pathloss}  
\end{figure}

Fig.~\ref{figure_SE_pathloss} shows the cumulative distribution functions (CDFs) of the SE per UE for different SNR realizations, for each of the channel models. The SNR variations lead to substantial differences in SE among the UEs. For all channel models, the UEs with low SNRs have a negligible gap between the case with correlated BS distortion and where the correlation is neglected. Clearly, it is not the BS distortion that limits the SE in these cases. When considering the UEs with high SE (i.e., strong SNR), the gap is wider and depends strongly on the channel model and combining scheme. For the Rayleigh fading cases in Figs.~\ref{figure_SE_pathloss}(a) and \ref{figure_SE_pathloss}(b), the curves have almost the same shape. When using DA-MMSE, neglecting the correlation leads to around 6\% higher SE than what is achievable, while the gap becomes as large as 25\% when using DA-MR. Moreover, the figure reaffirms that there is much to gain from using DA-MMSE when there is hardware distortion.

The curves have a different shape in the free-space propagation scenario, since the channel vectors are nearly orthogonal in this case \cite{massivemimobook}, except for UEs with very similar angles to the BS \cite{Ngo2014a}. DA-MR and DA-MMSE become  identical for the UEs with high SNR, since the distortion dominates and the BS distortion is hard to suppress in this case \cite{Mollen2018a}. If the distortion correlation is neglected, the SE becomes up to 10\% larger than in reality.

\begin{observation}
When there are SNR variations, the distortion correlation has negligible impact on the uplink SE for low-SNR UEs, while the impact is noticeable for high-SNR UEs. This observation applies to a variety of channel models.
\end{observation}

Note that when one UE has much higher SNR than the others, the vast majority of the distortion will be ``created'' by this UE's signal. Hence, we can expect the distortion to behave somewhere in between $K=1$ in Fig.~\ref{figure_SE_users} and the point where all the UEs have equal SNR.
In practice, there can be 50\,dB pathloss differences for UEs in a cell, but uplink power control is normally used to compress the differences (i.e., avoid near-far effects), leading to pathloss differences of the type considered in Fig.~\ref{figure_SE_pathloss}. If max-min fairness power-control is used, then the pathloss differences are removed completely.

\subsection{Impact of ADC Resolution}

\begin{figure}
  \centering  
    \includegraphics[width=0.5\textwidth]{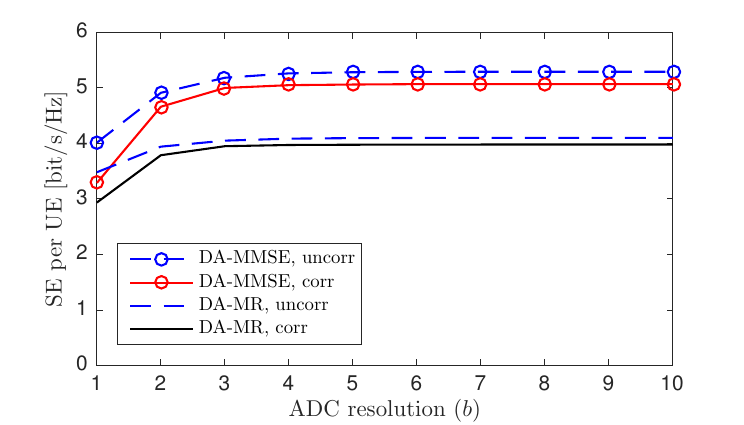} 
      \caption{SE per UE as a function of the ADC resolution $b$ with $K=5$ and $M=100$. We compare DA-MMSE and DA-MR when correlation of the BS distortion is either neglected ($\mathrm{uncorr}$) or accounted for ($\mathrm{corr}$).} \label{figure_SE_quantization}  
\end{figure}

We will now illustrate to what extent the quantization distortion and its correlation between the antennas impact the SE. 
The same basic setup as in Fig.~\ref{figure_SE_users} is considered. Fig.~\ref{figure_SE_quantization} shows the SE per UE for $K=5$ and the same parameter values as in Fig.~\ref{figure_SE_users}, except that the ADC resolution $b$ is now varied from 1 to 10 bits. The SE grows monotonically with $b$, but saturates after $b=4$. If there were SNR variations between the UEs, or fewer UEs, the saturation would occur at slightly larger bit resolutions. However, since today's systems have ADC resolutions of around 15 bits and there are 6\,bit ADCs that only consume 1.3\,mW while operating at 1\,Gsample/s \cite{Choo2016a}, it is unlikely that the quantization distortion will be a limiting factor in practical Massive MIMO systems. In fact, one might want to have even more bits in practice, to achieve robustness against unintentional interference (e.g., from the adjacent band \cite{Mollen2018a}) and intentional jamming.

\begin{observation}
Quantization distortion is not the main limiting factor for the SE in Massive MIMO, for practical ranges of ADC resolutions.
\end{observation}

\subsection{Asymptotic Analysis}

\begin{figure} 
        \centering
        \begin{subfigure}[b]{\columnwidth} \centering 
                \includegraphics[width=\textwidth]{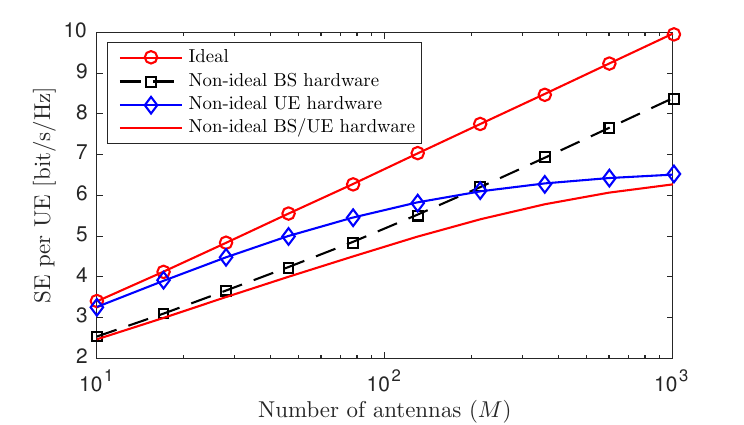} 
                \caption{DA-MMSE combining} 
                \label{figure_SE_asymptotics_MMSE}
        \end{subfigure} 
        \begin{subfigure}[b]{\columnwidth} \centering
                \includegraphics[width=\textwidth]{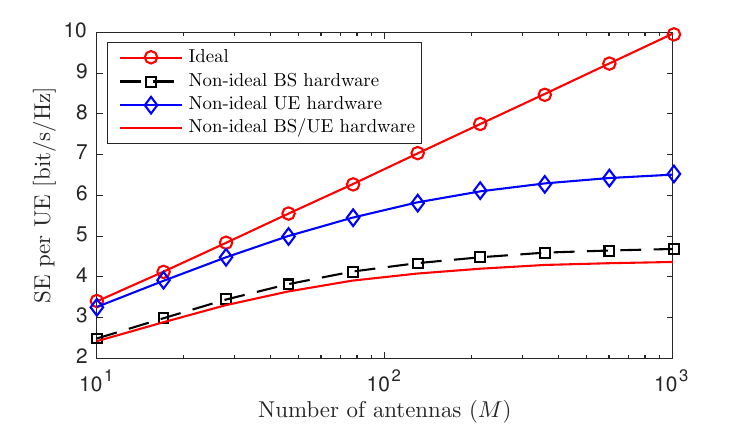} 
                \caption{DA-MR combining}  
                \label{figure_SE_asymptotics_MR} 
        \end{subfigure} 
        \caption{SE per UE as a function of $M$ when using either DA-MMSE or DA-MR combining. The asymptotic behaviors with either ideal or non-ideal hardware are evaluated.} 
        \label{figure_SE_asymptotics}  
\end{figure}

The distortion characteristics are important when analyzing the asymptotic SE when $M \to \infty$. To study this case, using the closed-form expressions developed in Section~\ref{sec:non-linearities}, we neglect the quantization distortion and consider $K=1$, which is the case when the distortion correlation is most influential (although it cannot be considered Massive MIMO due to the single UE antenna). Fig.~\ref{figure_SE_asymptotics} shows the SE for a varying number of antennas, reported in logarithmic scale from $M=10$ to $M=1000$, for an SNR of 0\,dB and i.i.d~Rayleigh fading. To study what type of hardware distortion limits the SE, we show the results with ideal hardware, non-ideal hardware at both the UE and BS (using the same parameters for non-linearities as above), and when only having non-ideal hardware at either the UE or the BS.

There is a substantial performance gap between having ideal hardware and the realistic case of non-ideal hardware at both the UE and BS. In fact, the gap grows as $\log_2(M)$ since the SE has no upper limit when using ideal hardware. The choice between DA-MMSE and DA-MR has a huge impact on the asymptotic limit under hardware impairments.
When using DA-MR, the convergence to the limit is fast and the curve with only non-ideal BS hardware gives a similar convergence, which implies that the BS distortion is the main limiting factor. 
In contrast, when using DA-MMSE, the curve with only non-ideal BS hardware goes to infinity, as expected from Section~\ref{subsec:distortion-directivtity}, where we demonstrated that spatial receiver processing can completely remove the BS distortion. Consequently, the curve with only non-ideal UE hardware converges to the same limit as the curve with non-ideal hardware at both the UE and BS. Note that it does not matter if the BS distortion is correlated or uncorrelated in this limit, but many hundreds of antennas are needed to fully neglect the BS distortion.

\begin{observation}
When using the optimal DA-MMSE combining, the UE distortion is the limiting factor as $M \to \infty$, since the BS distortion is suppressed by spatial processing. Since the suboptimal DA-MR combining does not suppress the BS distortion, it can become the main asymptotic limiting factor when using this scheme.
\end{observation}

This observation explains why different authors have reached different conclusions when studying the asymptotic SE under hardware impairments. For example,  \cite{Bjornson2014a,massivemimobook} claim that it is the UE distortion that limits the asymptotic SE and these works advocate for using combining schemes that suppresses interference and distortion. In contrast, \cite{Mollen2018a} notes that the BS distortion ``\emph{limits the effective SINR that can be achieved, even if the number of antennas is increased}'' since the paper assumes MR combining and does not consider UE distortion.

\subsection{Imperfect CSI}
\label{subsec:imperfect_CSI}

\begin{figure}
  \centering  
    \includegraphics[width=0.5\textwidth]{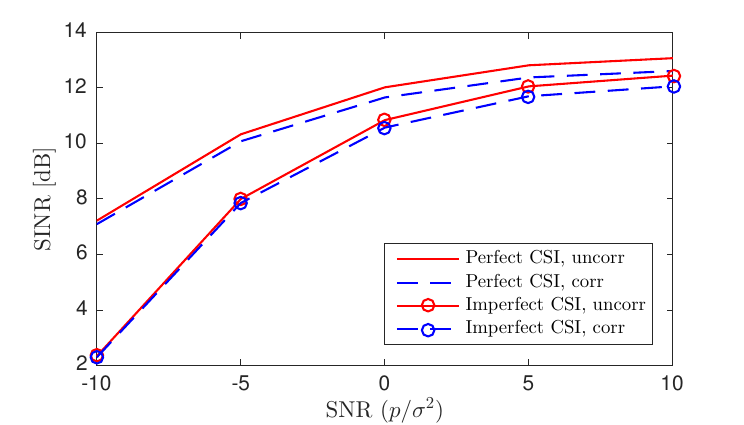} 
      \caption{SINR per UE as a function of the SNR with $K=5$ and $M=100$. We compare DA-MR with perfect and imperfect CSI in two cases: when correlation of the BS distortion is either neglected ($\mathrm{uncorr}$) or accounted for ($\mathrm{corr}$).} \label{figure_imperfectCSI}  
\end{figure}

Until now, we have assumed that the BS knows the channel $\vect{H}$ perfectly. The extension of our analytical results to imperfect CSI is non-trivial and deserves to be studied in detail in a separate paper. However, to demonstrate that nothing radically different is expected to happen, we will provide a numerical comparison between the performance achieved with perfect CSI and when the channels are estimated using least-square estimation. In the latter case, $K$-length orthogonal pilot sequences from a DFT matrix are transmitted in every coherence interval \cite[Sec.~3]{massivemimobook}. We consider the same setup as in Fig.~\ref{figure_SE_users} but with $K=5$ and varying SNR. 
Fig.~\ref{figure_imperfectCSI} shows the SINR per UE that is obtained by averaging over the channel realizations in the numerator and denominator of \eqref{SE-non-ideal-tx}. Note that if one would derive an achievable SE for the imperfect CSI case (which is a non-trivial task), it would probably not contain that exact SINR expression but something similar.
We only consider DA-MR since the extension of DA-MMSE combining to the imperfect CSI case is also non-trivial.

Fig.~\ref{figure_imperfectCSI} shows that the SINR loss from having imperfect CSI is substantial at low SNR (e.g., $-10$\,dB), while it is small at high SNR (e.g., $10$\,dB). The general behavior is otherwise the same in both cases: The BS distortion correlation has a negligible impact at low SNRs and a small but noticeable gap at higher SNRs. Hence, we expect all or most of the observations that are made in this paper to hold true also under imperfect CSI.

\section{Conclusions and Outlook}
\label{sec:conclusion}

The hardware distortion in a multiple-antenna BS is generally correlated across antennas. 
The correlation reduces the SINR, but we have shown that its impact on the SE is negligible, particularly when using DA-MMSE combining, which can utilize the correlation to effectively suppress the distortion.
Even in Massive MIMO systems with 100-200 antennas, approximating the BS distortion as uncorrelated when computing the SE only leads to overestimating the SE by a few percent and the bias reduces as more UEs are added. Since Massive MIMO is typically designed to serve tens of UEs, the error is negligible in the typical use cases. We demonstrated this by first deriving SE expressions with arbitrary quasi-memoryless distortion functions and then quantifying the impact of third-order AM-AM non-linearities and quantization errors. 
Analytical results were used to establish the basic phenomena qualitatively and then numerical results were used to quantify their impact on the SE.

The worst-case scenario for hardware distortion seems to be when serving one single-antenna UE at high-SNR in free-space propagation \cite{Mollen2018a}. In this special case, which is not MIMO since the UE has only one antenna, approximating the BS distortion as uncorrelated leads to an SE overestimation of around 1\,bit/s/Hz, assuming that the BS and UE have hardware of similar quality. This can be shown by comparing the SEs in the case where the BS distortion is negligible (e.g., uncorrelated distortion) with it being equally large as the UE distortion:
\begin{align}  \notag
&\underbrace{\log_2 \left(  1 + \frac{\kappa p M}{(1-\kappa)p M + O(1)}  \right)}_\textrm{BS distortion is negligible and included in $O(1)$} \\ &\quad- \underbrace{\log_2 \left(  1 + \frac{\kappa p M}{2 (1-\kappa)p M + O(1)}  \right)}_\textrm{BS distortion is equally large as UE distortion} \notag \\ & \quad \to \log_2\left( \frac{1}{1-\frac{\kappa}{2}} \right) < 1, \,\,\, M \to \infty
\end{align}
where $O(1)$ denotes the interference, distortion, and noise terms that are independent of $M$.
 In the typical Massive MIMO scenarios studied in this paper, the approximation error is substantially smaller than 1\,bit/s/Hz.

Additional phenomena may arise when using other channel propagation and hardware models as well as processing methods. In a multi-cell setup, the inter-cell interference reduces the SINR, in the same way as additional intra-cell UEs with low SNRs would do. If the interference increases, the relative impact of hardware distortion will reduce, making the uncorrelated-distortion approximation more accurate. Frequency-selective fading leads to reduced correlation \cite{Mollen2018b}, since every channel tap basically acts as an individual UE channel when computing the distortion characteristics. The inclusion of AM-PM distortion will likely also lower the correlation. The worst-case assumption of treating distortion as noise when computing the SE is practically convenient but might be vastly suboptimal. Compensation algorithms can be used to mitigate distortion in the digital baseband; particularly when dealing with non-destructive non-linearities that in principle can be inverted, although modeling inaccuracies and noise amplification will limit the invertibility in practice. For example, when using practical finite-sized constellations, we can apply the same receive combining as described in this paper and the same SINR is achievable, but a maximum a posteriori detector needs to be designed to adjust the decision boundaries to the distortion characteristics.

In conclusion, the uncorrelated distortion model advocated in \cite{Schenk2008a,Bjornson2014a,massivemimobook} (and used in numerous other papers) gives accurate results when analyzing the SE of multiuser MIMO systems. However, when using this model, one should always verify that the considered setup is one where the distortion indeed has negligible impact on the SE. 
We demonstrated that Massive MIMO is such a setup, for various numbers of UEs, channel models, and SNR ranges. Hence, although it might be ``{\emph{physically inaccurate}}'' to neglect distortion correlation, we can do it when analyzing the SE. The only reservation is that a BS deployed for Massive MIMO (e.g., $M\geq 100$, $K\geq 10$) will likely also perform single-user SIMO (single-input multiple-output) communication when the traffic is low and then the correlation is more influential, but the SE loss should not be more than 1\,bit/s/Hz.

\appendices

\section{Proof of Lemma~\ref{lemma:nonlinear_distortion_matrix}} \label{app:lemma:nonlinear_distortion_matrix}

The conditional correlation matrix of $\vect{z}$ has elements
\begin{align} \notag
& [\vect{C}_{zz} ]_{ij} = \EcondH\{ g_i(u_i) \left( g_j(u_j) \right)^*\} \\  & =  \EcondH\{ u_i u_j^*\} - a_i \EcondH\{  |u_i|^2 u_i  u_j^* \}- a_j \EcondH\{ u_i |u_j|^2 u_j^* \}  
 \notag \\  & \quad 
\label{eq:derivationCzz1} + a_i a_j \EcondH\{ |u_i|^2 |u_j|^2 u_i u_j^* \} \\ \notag
&=  \rho_{ij} \!-\! 2  a_i  \rho_{ii} \rho_{ij} \!-\! 2  a_j \rho_{jj} \rho_{ij}  \!+\! a_i a_j (2 | \rho_{ij}|^2 \rho_{ij} \!+\! 4 \rho_{ij} \rho_{ii} \rho_{jj} ) \notag \\
& = (1 - 2 a_i \rho_{ii}) \rho_{ij} (1 - 2 a_j \rho_{jj}) + 2 a_i a_j  | \rho_{ij}|^2  \rho_{ij} \label{eq:Czz}
 \end{align}
where the first expectation in \eqref{eq:derivationCzz1} equals $\rho_{ij}$, while the second and third expectations can be computed using Lemma~\ref{lemma:bussgang}. To compute the last expectation, we follow the procedure in the proof of Lemma~\ref{lemma:bussgang} to show that
$u_i = \frac{\rho_{ij}}{\rho_{jj}} u_j + \epsilon$, where $\epsilon \sim \CN (0,\rho_{ii}-{|\rho_{ij}|^2}/{\rho_{jj}} )$ is independent of $u_j$. Substituting this into the last expectation, it follows that
\begin{align} \notag
&\EcondH\{ |u_i|^2 |u_j|^2 u_i u_j^* \}  \\ \notag &
= \EcondH \left\{ \left|\frac{\rho_{ij}}{\rho_{jj}} u_j + \epsilon \right|^2 \! |u_j|^2 \left(\frac{\rho_{ij}}{\rho_{jj}} u_j + \epsilon\right) u_j^* \right\} \\ \notag
& = \left|\frac{\rho_{ij}}{\rho_{jj}}  \right|^2 \frac{\rho_{ij}}{\rho_{jj}} \EcondH \{ |u_j|^6  \} + 2 \frac{\rho_{ij}}{\rho_{jj}}  \EcondH \{|u_j|^4 \} \EcondH\{ |\epsilon |^2\} \\ \notag
& =  \left|\frac{\rho_{ij}}{\rho_{jj}}  \right|^2 \frac{\rho_{ij}}{\rho_{jj}} 6 \rho_{jj}^3+ 2 \frac{\rho_{ij}}{\rho_{jj}}  2 \rho_{jj}^2\left(\rho_{ii}-\frac{|\rho_{ij}|^2}{\rho_{jj}} \right) \\
& = 2 | \rho_{ij}|^2 \rho_{ij} + 4 \rho_{ij} \rho_{ii} \rho_{jj}.
\end{align}
By using \eqref{eq:DCssD} and \eqref{eq:Czz}, we obtain the elements of the distortion term's correlation matrix in \eqref{eq:Cetaeta} as
\begin{align} \label{eq:Cetaeta_expression1_proof}
 [\vect{C}_{\eta \eta} ]_{ij} =  [\vect{C}_{zz} ]_{ij} - [\vect{D} \vect{C}_{uu} \vect{D}^{\Htran}]_{ij} = 2 a_i a_j  | \rho_{ij}|^2  \rho_{ij}.
\end{align}
This is the final result given in \eqref{eq:Cetaeta_expression1}.

\section{Proof of Lemma~\ref{lemma:ratios-of-distortion}} \label{app:lemma:ratios-of-distortion}

For the assumed channel and hardware model, we have $\mathbb{E}\{ \| \vect{h}_k \|^2 \} = M$ and $a_m = \frac{\alpha}{\boff pK}$. Furthermore, using \eqref{eq:Cetaeta_expression} and $\rho_{ij} = p \sum_{n=1}^{K} h_{ni} h_{nj}^*$, we have
\begin{align} \notag
& \mathbb{E}\{ \vect{h}_k^{\Htran} \vect{C}_{\eta \eta} \vect{h}_k\} = 
\sum_{i=1}^{M} \sum_{j=1}^{M} \mathbb{E}\{ h_{ki}^* [\vect{C}_{\eta \eta}]_{ij} h_{kj} \}  \\
&= \sum_{i=1}^{M} \sum_{j=1}^{M}  2 a_i a_j p^3 \underbrace{\mathbb{E} \left\{ h_{ki}^* \left| \sum_{n=1}^{K} h_{ni} h_{nj}^* \right|^2  \sum_{l=1}^{K} h_{li} h_{lj}^* h_{kj} \right\}}_{=B_{ij}}
\end{align}
where $2 a_i a_j p^3 = 2 \frac{\alpha^2 p}{\boff^2 K^2}$. The expectations $B_{ij}$ can be computed by expanding the summations and then using the following lemma.

\begin{lemma} \label{lemma-moments} 
If $h \sim \CN(0,1)$, then for $p=1,2,\ldots$ we have $\mathbb{E}\{ |h|^{2p} \} = p!$.
\end{lemma}
\begin{IEEEproof} 
This follows from the moments of the exponential distribution, since
 $ |h|^{2} \sim \mathrm{Exp}(1)$.
\end{IEEEproof}

Using Lemma~\ref{lemma-moments} results in 
\begin{equation}
B_{ij}= \begin{cases}
K^3 +6K^2 + 11K + 6& i = j, \\
2K +2 & i \neq j.
\end{cases}
\end{equation} 
Since there are $M$ terms with $i=j$ and $M(M-1)$ terms with $i \neq j$, we finally obtain \eqref{eq:ratio-of-distortion} after some algebra. When the correlation between distortion terms is neglected, \eqref{eq:ratio-of-distortion-uncorr} is achieved analogously by setting $B_{ij}=0$ for $i \neq j$.

\section{Proof of Lemma~\ref{lemma:UE-distortion-term}} \label{app:lemma:UE-distortion-term}

For the assumed channel and hardware model, we have $a_m = \frac{\alpha}{\boff pK}$, and $\rho_{mm} = p \sum_{n=1}^{K} | h_{nm} |^2$. Moreover, 
\begin{align} \notag
&\mathbb{E}\{ | \vect{h}_k^{\Htran} \vect{D} \vect{h}_k |^2\} = 
\mathbb{E} \left\{ \left| \sum_{m=1}^{M} |h_{km}|^2(1 - 2 a_m \rho_{mm}) \right|^2 \right\} \\ \notag &
= \mathbb{E} \left\{ \left| \sum_{m=1}^{M} |h_{km}|^2 \left(1 - A \sum_{n=1}^{K} | h_{nm} |^2 \right) \right|^2 \right\} \\  \label{eq:UE-dist-deriv1}
&= \sum_{m_1=1}^{M} 
\sum_{m_2=1}^{M} \Bigg( \mathbb{E} \left\{ |h_{km_1}|^2  |h_{km_2}|^2 \right\} \\  \label{eq:UE-dist-deriv2}
& - 2 A  \, \mathbb{E} \left\{   |h_{km_1}|^2  | h_{km_2}|^2 \sum_{n=1}^{K} | h_{nm_1} |^2 \right\} \\  \label{eq:UE-dist-deriv3} &+ A^2 \,  \mathbb{E} \left\{ |h_{km_1}|^2  |h_{km_2}|^2  \left( \sum_{n=1}^{K} | h_{nm_1} |^2 \right) \!\! \left( \sum_{n=1}^{K} | h_{nm_2} |^2 \right)
 \right\} \! \Bigg)
\end{align}
where $A = \frac{2\alpha}{\boff K}$. It remains to compute the expectations in \eqref{eq:UE-dist-deriv1}--\eqref{eq:UE-dist-deriv3} and divide with $\mathbb{E}\{ \| \vect{h}_k \|^2 \} = M$ to obtain \eqref{eq:UE-distortion-term}. Direct computation of \eqref{eq:UE-dist-deriv1} using Lemma~\ref{lemma-moments} yields $M^2+M$, while the expression in \eqref{eq:UE-dist-deriv2} becomes
\begin{align} \notag
&-2 A \left( (K-1) (M^2+M) + 2M(M-1) + 6M \right) \\ &= -2 A M (MK + K+M+3).
\end{align}
Finally, \eqref{eq:UE-dist-deriv3} is computed by expanding all the summations and identifying the correlated terms. This results in
\begin{equation}
A^2 M \left(MK^2+8K+11+2MK+K^2+M \right).
\end{equation}

\section*{Acknowledgment}

The authors would like to thank Prof.~Erik G.~Larsson for useful feedback on our manuscript.

\bibliographystyle{IEEEtran}
% argument is your BibTeX string definitions and bibliography database(s)
\bibliography{IEEEabrv,refs.bib}

% Generated by IEEEtran.bst, version: 1.14 (2015/08/26)
\begin{thebibliography}{10}
\providecommand{\url}[1]{#1}
\csname url@samestyle\endcsname
\providecommand{\newblock}{\relax}
\providecommand{\bibinfo}[2]{#2}
\providecommand{\BIBentrySTDinterwordspacing}{\spaceskip=0pt\relax}
\providecommand{\BIBentryALTinterwordstretchfactor}{4}
\providecommand{\BIBentryALTinterwordspacing}{\spaceskip=\fontdimen2\font plus
\BIBentryALTinterwordstretchfactor\fontdimen3\font minus
  \fontdimen4\font\relax}
\providecommand{\BIBforeignlanguage}[2]{{%
\expandafter\ifx\csname l@#1\endcsname\relax
\typeout{** WARNING: IEEEtran.bst: No hyphenation pattern has been}%
\typeout{** loaded for the language `#1'. Using the pattern for}%
\typeout{** the default language instead.}%
\else
\language=\csname l@#1\endcsname
\fi
#2}}
\providecommand{\BIBdecl}{\relax}
\BIBdecl

\bibitem{Bjornson2018a}
E.~Bj\"{o}rnson, L.~Sanguinetti, and J.~Hoydis, ``Can hardware distortion
  correlation be neglected when analyzing uplink {SE} in {Massive} {MIMO}?'' in
  \emph{Proc.~IEEE SPAWC}, 2018.

\bibitem{Schenk2008a}
T.~Schenk, \emph{{RF} imperfections in high-rate wireless systems: Impact and
  digital compensation}.\hskip 1em plus 0.5em minus 0.4em\relax Springer, 2008.

\bibitem{Bussgang1952a}
J.~J. Bussgang, ``Crosscorrelation functions of amplitude-distorted {Gaussian}
  signals,'' Research Laboratory of Electronics, Massachusetts Institute of
  Technology, Tech. Rep. 216, 1952.

\bibitem{Fletcher2007a}
A.~K. Fletcher, S.~Rangan, V.~K. Goyal, and K.~Ramchandran, ``Robust predictive
  quantization: Analysis and design via convex optimization,'' \emph{{IEEE} J.
  Sel. Topics Signal Process.}, vol.~1, no.~4, pp. 618--632, 2007.

\bibitem{Hassibi2003a}
B.~Hassibi and B.~M. Hochwald, ``How much training is needed in
  multiple-antenna wireless links?'' \emph{{IEEE} Trans. Inf. Theory}, vol.~49,
  no.~4, pp. 951--963, 2003.

\bibitem{Parkvall2017a}
S.~Parkvall, E.~Dahlman, A.~Furusk\"ar, and M.~Frenne, ``{NR}: The new {5G}
  radio access technology,'' \emph{IEEE Communications Standards Magazine},
  vol.~1, no.~4, pp. 24--30, 2017.

\bibitem{Larsson2014a}
E.~G. Larsson, F.~Tufvesson, O.~Edfors, and T.~L. Marzetta, ``Massive {MIMO}
  for next generation wireless systems,'' \emph{{IEEE} Commun. Mag.}, vol.~52,
  no.~2, pp. 186--195, 2014.

\bibitem{massivemimobook}
E.~Bj\"{o}rnson, J.~Hoydis, and L.~Sanguinetti, ``Massive {MIMO} networks:
  {Spectral}, energy, and hardware efficiency,'' \emph{Foundations and
  Trends{\textregistered} in Signal Processing}, vol.~11, no. 3-4, pp.
  154--655, 2017.

\bibitem{Marzetta2010a}
T.~L. Marzetta, ``Noncooperative cellular wireless with unlimited numbers of
  base station antennas,'' \emph{{IEEE} Trans. Wireless Commun.}, vol.~9,
  no.~11, pp. 3590--3600, 2010.

\bibitem{Bjornson2016a}
E.~Bj{\"{o}}rnson, E.~G. Larsson, and M.~Debbah, ``Massive {MIMO} for maximal
  spectral efficiency: How many users and pilots should be allocated?''
  \emph{{IEEE} Trans. Wireless Commun.}, vol.~15, no.~2, pp. 1293--1308, 2016.

\bibitem{ashikhmin2012pilot}
A.~Ashikhmin and T.~Marzetta, ``Pilot contamination precoding in multi-cell
  large scale antenna systems,'' in \emph{IEEE International Symposium on
  Information Theory Proceedings (ISIT)}, 2012, pp. 1137--1141.

\bibitem{Yin2013a}
H.~Yin, D.~Gesbert, M.~Filippou, and Y.~Liu, ``A coordinated approach to
  channel estimation in large-scale multiple-antenna systems,'' \emph{{IEEE} J.
  Sel. Areas Commun.}, vol.~31, no.~2, pp. 264--273, 2013.

\bibitem{BHS18A}
E.~Bj{\"o}rnson, J.~Hoydis, and L.~Sanguinetti, ``Massive {MIMO} has unlimited
  capacity,'' \emph{IEEE Transactions on Wireless Communications}, vol.~17,
  no.~1, pp. 574--590, 2018.

\bibitem{Shepard2013a}
C.~Shepard, H.~Yu, and L.~Zhong, ``Argosv2: A flexible many-antenna research
  platform,'' in \emph{Proc.~ACM MobiCom}, 2013.

\bibitem{Harris2017a}
P.~Harris, S.~Malkowsky, J.~Vieira, E.~Bengtsson, F.~Tufvesson, W.~B. Hasan,
  L.~Liu, M.~Beach, S.~Armour, and O.~Edfors, ``Performance characterization of
  a real-time {Massive MIMO} system with {LOS} mobile channels,'' \emph{IEEE
  Journal on Selected Areas in Communications}, vol.~35, no.~6, pp. 1244--1253,
  June 2017.

\bibitem{Bai2013a}
Q.~Bai, A.~Mezghani, and J.~A. Nossek, ``On the optimization of {ADC}
  resolution in multi-antenna systems,'' in \emph{Proc.~IEEE ISWCS}, 2013.

\bibitem{Orhan2015a}
O.~Orhan, E.~Erkip, and S.~Rangan, ``Low power analog-to-digital conversion in
  millimeter wave systems: Impact of resolution and bandwidth on performance,''
  in \emph{Proc.~IEEE ITA}, 2015.

\bibitem{Bjornson2014a}
E.~Bj{\"{o}}rnson, J.~Hoydis, M.~Kountouris, and M.~Debbah, ``Massive {MIMO}
  systems with non-ideal hardware: Energy efficiency, estimation, and capacity
  limits,'' \emph{{IEEE} Trans. Inf. Theory}, vol.~60, no.~11, pp. 7112--7139,
  2014.

\bibitem{Jacobsson2017b}
S.~Jacobsson, G.~Durisi, M.~Coldrey, U.~Gustavsson, and C.~Studer, ``Throughput
  analysis of massive {MIMO} uplink with low-resolution {ADCs},'' \emph{{IEEE}
  Trans. Wireless Commun.}, vol.~16, no.~6, pp. 4038--4051, 2017.

\bibitem{Moghadam2012a}
N.~N. Moghadam, P.~Zetterberg, P.~H\"{a}ndel, and H.~Hjalmarsson, ``Correlation
  of distortion noise between the branches of {MIMO} transmit antennas,'' in
  \emph{Proc.~IEEE PIMRC}, 2012.

\bibitem{Mollen2018b}
C.~Moll\'{e}n, U.~Gustavsson, T.~Eriksson, and E.~G. Larsson, ``Spatial
  characteristics of distortion radiated from antenna arrays with transceiver
  nonlinearities,'' \emph{{IEEE} Trans. Wireless Commun.}, vol.~17, no.~10, pp.
  6663--6679, 2018.

\bibitem{Gustavsson2014a}
U.~Gustavsson, C.~Sanch\'ez-Perez, T.~Eriksson, F.~Athley, G.~Durisi,
  P.~Landin, K.~Hausmair, C.~Fager, and L.~Svensson, ``On the impact of
  hardware impairments on massive {MIMO},'' in \emph{Proc.~IEEE GLOBECOM},
  2014.

\bibitem{Handel2018a}
P.~H\"andel and D.~R\"onnow, ``Dirty {MIMO} transmitters: Does it matter?''
  \emph{{IEEE} Trans. Wireless Commun.}, vol.~17, no.~8, pp. 5425--5436, 2018.

\bibitem{Zou2015a}
Y.~Zou, O.~Raeesi, L.~Antilla, A.~Hakkarainen, J.~Vieira, F.~Tufvesson, Q.~Cui,
  and M.~Valkama, ``Impact of power amplifier nonlinearities in multi-user
  massive {MIMO} downlink,'' in \emph{Proc.~IEEE~GLOBECOM Workshops}, 2015.

\bibitem{Larsson2018a}
E.~G. Larsson and L.~V. der Perre, ``Out-of-band radiation from antenna arrays
  clarified,'' \emph{{IEEE} Commun. Lett.}, vol.~7, no.~4, pp. 610--613, 2018.

\bibitem{Mollen2018a}
C.~Moll\'{e}n, U.~Gustavsson, T.~Eriksson, and E.~G. Larsson, ``Impact of
  spatial filtering on distortion from low-noise amplifiers in massive {MIMO}
  base stations,'' \emph{{IEEE} Trans. Commun.}, 2018, to appear.

\bibitem{Raich2002a}
R.~Raich and G.~Zhou, ``On the modeling of memory nonlinear effects of power
  amplifiers for communication applications,'' in \emph{Proc.~IEEE DSP
  Workshop}, 2002.

\bibitem{Kay1993a}
S.~M. Kay, \emph{Fundamentals of statistical signal processing: Estimation
  theory}.\hskip 1em plus 0.5em minus 0.4em\relax Prentice Hall, 1993.

\bibitem{Jacobsson2017a}
S.~Jacobsson, G.~Durisi, M.~Coldrey, T.~Goldstein, and C.~Studer, ``Quantized
  precoding for massive {MU-MIMO},'' \emph{{IEEE} Trans. Commun.}, vol.~65,
  no.~11, pp. 4670--4684, 2017.

\bibitem{Bjornson2015b}
E.~Bj{\"{o}}rnson, M.~Matthaiou, and M.~Debbah, ``Massive {MIMO} with non-ideal
  arbitrary arrays: Hardware scaling laws and circuit-aware design,''
  \emph{{IEEE} Trans. Wireless Commun.}, vol.~14, no.~8, pp. 4353--4368, 2015.

\bibitem{3GPP_PA_models}
Ericsson, ``Further elaboration on {PA} models for {NR},'' 3GPP TSG-RAN WG4,
  R4-165901, Tech. Rep., Aug. 2016.

\bibitem{Gao2015b}
X.~Gao, O.~Edfors, F.~Tufvesson, and E.~G. Larsson, ``Massive {MIMO} in real
  propagation environments: Do all antennas contribute equally?'' \emph{{IEEE}
  Trans. Commun.}, vol.~63, no.~11, pp. 3917--3928, 2015.

\bibitem{Lloyd1982a}
S.~Lloyd, ``Least squares quantization in {PCM},'' \emph{{IEEE} Trans. Inf.
  Theory}, vol.~28, no.~2, pp. 129--137, 1982.

\bibitem{Chan2015a}
C.-H. Chan, Y.~Zhu, S.-W. Sin, U.~Seng-Pan, and R.~P. Martins, ``{A 5.5mW 6b
  5GS/S 4x-interleaved 3b/cycle SAR ADC in 65nm CMOS},'' in \emph{Proc.~IEEE
  ISSCC}, 2015.

\bibitem{Choo2016a}
K.~D. Choo, J.~Bell, and M.~P. Flynn, ``Area-efficient {1GS/s 6b SAR ADC} with
  charge-injection-cell-based {DAC},'' in \emph{Proc.~IEEE ISSCC}, 2016.

\bibitem{Studer2016a}
C.~Studer and G.~Durisi, ``Quantized massive {MU-MIMO-OFDM} uplink,''
  \emph{{IEEE} Trans. Commun.}, vol.~64, no.~6, pp. 2387--2399, 2016.

\bibitem{Mollen2017a}
C.~Moll\'{e}n, J.~Choi, E.~G. Larsson, and R.~W. Heath, ``Achievable uplink
  rates for massive {MIMO} with coarse quantization,'' in \emph{Proc.~IEEE
  ICASSP}, 2017.

\bibitem{Verenzuela2017b}
D.~Verenzuela, E.~Bj\"{o}rnson, and M.~Matthaiou, ``Per-antenna hardware
  optimization and mixed resolution {ADCs} in uplink massive {MIMO},'' in
  \emph{Proc. Asilomar}, 2017.

\bibitem{Ngo2014a}
H.~Q. Ngo, E.~G. Larsson, and T.~L. Marzetta, ``Aspects of favorable
  propagation in massive {MIMO},'' in \emph{Proc.~EUSIPCO}, 2014.

\end{thebibliography}
%

% that's all folks
\end{document}